\documentclass[english,12pt,onecolumn]{IEEEtran}

\usepackage{cite}
\usepackage{amsmath,amssymb,amsfonts}
\usepackage{algorithmic}
\usepackage{graphicx}
\usepackage{algorithm}
\usepackage{hyperref}
\usepackage{textcomp}

\usepackage{bm, mathtools, marvosym, accents}
\usepackage{epstopdf}
\usepackage{tikz, environ, float, xcolor}
\usetikzlibrary{shapes, arrows, calc, patterns, decorations.pathreplacing, backgrounds, positioning, fit}
\usepackage{enumitem} 

\usepackage{setspace}
\newtheorem{theorem}{Theorem}
\newtheorem{lemma}{Lemma}
\newtheorem{corollary}{Corollary}
\newtheorem{assumption}{Assumption}

\newtheorem{example}{Example}
\newtheorem{remark}{Remark}
\newtheorem{proposition}{Proposition}

\newcommand{\ubar}[1]{\underaccent{\bar}{#1}}

\DeclareMathOperator{\sgn}{sgn}
\DeclareMathOperator{\id}{Id}

\DeclareMathOperator{\diag}{diag}

\DeclareMathOperator{\acos}{acos}

\newcommand{\argmin}{\operatornamewithlimits{argmin}}

\begin{document}
% 2024.02.10 for the draft 
% \doublespacing

\title{Constructive Safety Control}
\author{Si Wu, Tengfei Liu, and Zhong-Ping Jiang, % <-this % stops a space
\thanks{This work was supported in part by National Key R \& D Program of China (2022ZD0119901), in part by Science and Technology Projects of Liaoning Province (2022JH2/101300238), in part by Liaoning Revitalization Talents Program XLYC2203076, in part by NSFC grants 62325303 and 62333004, and in part by NSF grant ECCS-2210320.}% <-this % stops a space
\thanks{S. Wu and T. Liu are with the State Key Laboratory of Synthetical Automation for Process Industries, Northeastern University, Shenyang, 110004, China {\tt\small(e-mails:
wusixstx@163.com; 2010306@stu.neu.edu.cn; tfliu@mail.neu.edu.cn)} }%
\thanks{Z. P. Jiang is with the Department of Electrical and Computer Engineering, New York University, 360 Jay Street, Brooklyn, NY 11201, USA {\tt\small (e-mail: zjiang@nyu.edu)}}%
\thanks{Corresponding author: T. Liu.}}

\maketitle

\begin{abstract}
This paper proposes a constructive approach to safety control of nonlinear cascade systems subject to multiple state constraints. New design ingredients include a unified characterization of safety and stability for systematic designs of safety controllers, and a novel technique of reshaping the feasible sets of quadratically constrained quadratic programming induced from safety control. The proposed method guarantees Lipschitz continuity of virtual control laws, enabling a stepwise constructive design. A refined nonlinear small-gain synthesis is employed to address the nonlinear uncertain interconnections between the resulting subsystems corresponding to different virtual control laws, and to guarantee the achievement of the safety control objective. When the safety constraints are removed, the proposed approach coincides with the standard constructive nonlinear control. The proposed safety-control algorithm is experimentally validated in a testbed involving a vertical takeoff and landing (VTOL) vehicle taking off in narrow spaces.
\end{abstract}


\section{Introduction}
\label{section_introduction}

% 1
\IEEEPARstart{A}{ttaining} primary objectives while ensuring safety constraints is vital for a wide range of engineering systems \cite{Ames-Xu-Grizzle-Tabuada-TAC-2017}, with promising applications including humanoid robots \cite{Escande-Mansard-Wieber-IJRR-2014}, manipulators \cite{Singletary-Nilsson-Gurriet-Ames-2019-IROS}, human-robot interaction systems \cite{Angeliki-Ioannis-Antonios-Ioannis-2020-SS}, and multi-agent systems \cite{Wang-Ames-Egerstedt-TRO-2017}. Tremendous efforts have been devoted to solving safety control problems of complex systems. Representative methods include trajectory planning and tracking \cite{Chen-Herbert-Hu-Pu-Fisac-Bansal-Han-Tomlin-2021-TAC}, reachability analysis \cite{Bansal-Chen-Herbert-Tomlin-2017-CDC}, the velocity obstacle approach \cite{Berg-Guy-Lin-Manocha11-RR-2011, Fiorini-Shiller-IJRR-1998}, contraction theory \cite{Singh-Majumdar-Slotine-Pavone-2017-ICRA}, to name only a few.

Among various approaches, the barrier-function-based method and its extensions, including zeroing, reciprocal, exponential, and adaptive barrier functions \cite{Ames-Coogan-Egerstedt-ECC-2019, Xiao-Belta-Cassandras-2022-TAC}, provide formal guarantees of system safety with enhanced robustness and flexibility. These methods have been shown extremely useful to solve safety control problems for systems with high-relative-degrees \cite{Nguyen-Sreenath-2016-ACC, Tan-Cortez-Dimarogonas-2022-TAC, Xiao-Belta-2022-TAC, Breeden-Panagou-Automatica-2023}, hybrid systems \cite{Nguyen-Sreenath-ACC-2016}, and nonlinear uncertain systems \cite{Xu-Tabuada-Grizzle-Ames-IFAC-2015, Jankovic-Auto-2018, Cosner-Singletary-Taylor-Molnar-Bouman-Ames-IROS-2021,Alan-Taylor-He-Ames-Orosz-TCST-2023}. Moreover, control barrier functions (CBF) allow seamless integration with other control methods, such as model predictive control \cite{Grandia-Taylor-Ames-Hutter-2021-ICRA}, adaptive control \cite{Nguyen-Sreenath-2022-ACC} and inverse optimal control \cite{Krstic-TAC-2023}. 

Quadratic programming (QP) is a powerful tool for real-time synthesis of controllers by incorporating different specifications simultaneously \cite{Nakamura-Book-1990, Escande-Mansard-Wieber-IJRR-2014, Mellinger-Kumar-ICRA-2011, Ames-Powell-CPS-2013}. For a system subject to safety constraints, a QP algorithm calculates the admissible control input that fulfills the constraints and minimally invades the nominal control inputs for primary objectives \cite{Ames-Xu-Grizzle-Tabuada-TAC-2017}.
The integration of barrier functions and QP algorithms has found various applications, including automotive safety \cite{Ames-Xu-Grizzle-Tabuada-TAC-2017}, robotic locomotion and manipulation \cite{Morris-Powell-Ames-CDC-2015, Cortez-Oetomo-Manzie-Choong-TCST-2019}, multi-robot systems \cite{Wang-Ames-Egerstedt-TRO-2017, Glotfelter-Cortes-Egerstedt-TAC-2020, Wilson-Egerstedt-CSM-2020}.

Aligning with engineering practice principles, constructive nonlinear control methods take advantage of structural properties to design controllers for nonlinear uncertain systems.
In particular, recursive design methods including backstepping and forwarding have been developed for benchmark systems in lower-triangular, upper-triangular and generalized forms \cite{Karafyllis-Jiang-Book-2011, Kokotovic-Arcak-Auto-2001, Krstic-Kokotovic-Kanellakopoulos-Book-1995, Marino-Tomei-Book-1996, Freeman-Kokotovic-Book-2008, Sepulchre-Jankovic-Kokotovic-Book-1997, Krstic-Deng-Book-1998, Dawson-Hu-Burg-Book-2019, Isidori-Book-1999, Mazenc-Praly-TAC-1996, Teel-Book-1993}. % 引用书，再加一些？Marino and Tomei (1995), Freeman and KokotovicH (1996b),  Sepulchre et al. (1997), KrsticH and Deng (1998), Dawson, Hu, and Burg (1998), and Isidori (1999) JankovicH, Sepulchre, and KokotovicH (1996替换), Teel (1992), Mazenc and Praly (1996)
These methods enable simultaneous controller design and stability analysis in a stepwise and recursive manner. Recently, backstepping designs have been extended to systems subject to safety constraints \cite{Taylor-Pio-Molnar-Ames-CDC-2022}. However, when it comes to systems subject to multiple safety constraints, there appears to be a lack of systematic results on constructive safety control. For safety control of lower-triangular systems, one major challenge is caused by the possibly non-Lipschitz, virtual control laws and the non-robustness of the controlled subsystems if multiple safety constraints are directly integrated by some optimization-based algorithms.

This paper takes a step forward toward solving the safety control problem for nonlinear cascade systems working in environments involving multiple obstacles.
Current relevant results include those based on exponential control barrier functions \cite{Nguyen-Sreenath-2016-ACC}, high-relative-degree control barrier functions \cite{Tan-Cortez-Dimarogonas-2022-TAC, Xiao-Belta-2022-TAC}, or backstepping control barrier functions \cite{Hsu-Xu-Ames-2015-ACC, Taylor-Pio-Molnar-Ames-CDC-2022}, mainly focusing on one single safety constraint. In the scenarios with more than one safety constraints, the optimization solution with safety constraints incorporated may become infeasible or non-Lipschitz with respect to the state of the plant \cite{Hager-SIAMControl-1979, Morris-Powell-Ames-CDC-2015, Isaly-Ghanbarpour-Sanfelice-Dixon-TAC-2024}. The non-Lipschitz continuity may lead to loss of robustness with respect to the response of the higher-order dynamics, compromising the safety of the closed-loop system \cite{Wu-Liu-Egerstedt-Jiang-2023-TAC}.

% 4. 
This paper introduces a novel approach to safety control of nonlinear cascade systems. We expect to design a safety controller tailored for nonlinear systems characterized by higher relative degrees, while also subject to multiple distinct safety constraints. First, we propose a safety controller based on quadratically constrained quadratic programming (QCQP) with guaranteed feasibility and input-to-state practical safety (ISpSf) for a class of nonlinear uncertain plants in the control-affine form subject to multiple constraints. Then, we contribute a feasible-set reshaping technique for QCQP algorithms, which projects the quadratic constraints involving uncertainties onto a predefined positive basis to address the non-Lipschitz issue. Finally, we solve the safety control problem for nonlinear cascade systems through a recursive design procedure and a small-gain synthesis. The basic idea is to transform the closed-loop system into an interconnected system and appropriately choose the interconnection gains to ensure safety. This paper is organized as follows:
\begin{itemize}
\item A QCQP-based safety control method for nonlinear uncertain plants with relative-degree-one is given in Section \ref{section_relative_degree_one};
\item A feasible-set reshaping technique for safety-oriented QCQP algorithms is proposed in Section \ref{section_feasible_set_reshaping};
\item A constructive approach to safety control of nonlinear cascade systems is developed in Section \ref{section_main_result}.
\end{itemize}
An experiment of a vertical takeoff and landing (VTOL) vehicle taking off in narrow spaces is employed to verify the proposed methods in Section \ref{section_experiment}.

% Compared with our previous work on safety control, this paper expands the scope of safety control to nonlinear plants with higher relative degrees. Additionally, it presents a condition that ensures the feasibility of the feasible set without relying on relaxation parameters. Furthermore, the paper introduces a lemma guiding the feasible-set reshaping technique. 

\section*{Notations and Preliminaries}

In this paper, $|\cdot|$ represents the Euclidean norm for real vectors and the induced $2$-norm for real matrices. For a real matrix $A$, we use $[A]_{i,j}$, $[A]_{i,:}$ and $[A]_{:, j}$ to represent the element at row $i$ and column $j$, the $i$-th row vector and the $j$-th column vector, respectively, and use $\min\{A\}$ and $\max\{A\}$ to represent the column vectors containing the minimum values and maximum values of the corresponding rows, respectively. 
We use $\odot$ to represent the Hadamard product: for $A\in\mathbb{R}^{m\times n}$ and $B\in\mathbb{R}^{1\times n}$, $C = A\odot B$ is defined by $[C]_{i,j}=[A]_{i,j}[B]_{1,j}$. For symmetric matrix $A\in\mathbb{R}^{n\times n}$, $\lambda_{\max}(A)$ and $\lambda_{\min}(A)$ take the maximum eigenvalue and the minimum eigenvalue, respectively.

A function $\alpha:\mathbb{R}_+\rightarrow\mathbb{R}_+$ is said to be of class $\mathcal{K}$, denoted by $\alpha\in\mathcal{K}$, if it is continuous and strictly increasing, and $\alpha(0)=0$; it is said to be of class $\mathcal{K}_{\infty}$, denoted by $\alpha\in\mathcal{K}_{\infty}$, if it is of class $\mathcal{K}$ and radially unbounded. A continuous function $\alpha:(-a, b)\rightarrow (-c,\infty)$ with constants $a,b,c\in\mathbb{R}_+\cup\{\infty\}$ is said to be of class $\mathcal{K}^e$, denoted by $\alpha\in\mathcal{K}^e$, if it is continuous and strictly increasing and $\alpha(0)=0$. 

A continuous function $\beta:\mathbb{R}_+\times\mathbb{R}_+\rightarrow\mathbb{R}_+$ is said to be of class $\mathcal{KL}$, denoted by $\beta\in\mathcal{KL}$, if, for each fixed $t\in\mathbb{R}_+$, $\beta(\cdot,t)$ is a class $\mathcal{K}$ function and, for each fixed $s>0$, $\beta(s,\cdot)$ is a decreasing function and satisfies $\lim_{t\rightarrow\infty}\beta(s,t)=0$. We use $\sgn(r)$ to represent the sign of $r\in\mathbb{R}$.

\section{Problem Formulation}
\label{section_main_result_problem_formulation}

This paper aims to address the difficulty caused by multiple safety constraints and uncertain dynamics in the safety control of nonlinear cascade systems. This section first introduces the class of systems to be studied in this paper and formulates the safety control problem. The study is still nontrivial even when the plant is of relative-degree-one. 

\subsection{Safety Control of Nonlinear Cascade Systems}
\label{section_problem_formulation_cascade_systems}

The study in this paper follows the convention of constructive nonlinear control, and focuses on nonlinear plants in the cascade form:
\begin{subequations} \label{eq_plant_dynamics}
\begin{align}
\dot{x}_i &= f_i(z_i)+(g_i+\delta_i(z_i))x_{i+1}, i=1,\ldots,m-1 \label{eq_dynamics_x_i} \\
\dot{x}_m &= f_m(z_m)+(g_m+\delta_m(z_m))u, \label{eq_dynamics_x_m}
\end{align}
\end{subequations}
where $x_i\in\mathbb{R}^{n_{x_i}}$ for $i=1,\ldots,m$ form the state, $z_i = [x_1;\ldots;x_i]\in \mathbb{R}^{n_{z_i}}$ with $n_{z_i}:=\sum_{j=1}^{i} n_{x_j}$, $u\in\mathbb{R}^{n_u}$ is the control input, $g_i\in\mathbb{R}^{n_{x_i}\times n_{x_{i+1}}}$ is a constant matrix, $f_i:\mathbb{R}^{n_{z_i}}\rightarrow\mathbb{R}^{n_{x_i}}$ and $\delta_i:\mathbb{R}^{n_{z_i}}\rightarrow\mathbb{R}^{n_{x_i}\times n_{x_{i+1}}}$ are locally Lipschitz functions with $n_{x_{m+1}}=n_u$. The first state $x_1$ is considered as the output.

Given an initial nominal controller 
\begin{align}
u_0=\rho_0(x_1), \label{eq_def_nominal_controller_highorder}
\end{align}
with $u_0\in\mathbb{R}^{n_{x_2}}$ representing the nominal control input and $\rho_0:\mathbb{R}^{n_{x_1}}\rightarrow\mathbb{R}^{n_{x_2}}$, and safety certificate functions $V_j:\mathcal{X}\rightarrow\mathbb{R}_+$ for $j=1,\ldots,n_c$ with $\mathcal{X}\subseteq\mathbb{R}^{n_{x_1}}$ being a nonempty open set, we expect to design a continuous safety controller in the following form:
\begin{align}
u = \rho(z_m),
\end{align}
where, following the convention of constructive nonlinear control, $\rho:\mathbb{R}^{n_{z_m}}\rightarrow\mathbb{R}^{n_u}$ is supposed to be in the recursive form
\begin{subequations} \label{eq_controller_high}
\begin{align}
\rho(z_m) &= x^*_{m+1}, \\
x^*_{i+1} &= \rho_i(x_i-x^*_i),~i=2,\ldots,m,~~~x^*_2 = \rho_1(x_1),
\end{align}
\end{subequations}
with $x^*_{i+1}\in\mathbb{R}^{n_{x_{i+1}}}$ being virtual control inputs and $\rho_i:\mathbb{R}^{n_{x_i}}\rightarrow\mathbb{R}^{n_{x_{i+1}}}$ for $i=1,\ldots,m$. Moreover, $\rho_1$ will be chosen based on $\rho_0$, such that the virtual control input $x^*_2$ minimally invades the nominal control input $u_0$. 

With the proposed safety controller, the closed-loop system composed of \eqref{eq_plant_dynamics} and \eqref{eq_controller_high} is expected to be safe in the sense that the state trajectory $x_1(t)$ keeps inside $\mathcal{X}$ whenever it is defined, and there exist $\breve{\beta}, \breve{\beta}_2,\ldots,\breve{\beta}_m \in \mathcal{KL}$ and constants $\breve{c},\breve{c}_2,\ldots,\breve{c}_m\ge 0$ such that for any initial state $x_1(0) \in \mathcal{X}$ and $x_i(0) \in \mathbb{R}^{n_{x_i}}$ for $i=2,\ldots,m$, 
\begin{subequations}\label{eq_objective_highorder}
\begin{align}
\breve{V}(x_1(t)) \le \max\{\breve{\beta}(&\max\{\breve{V}(x_1(0)), |\rho_0(x_1(0))|, \notag \\
&|x_2(0)|,\ldots,|x_m(0)|\},t),~ \breve{c}\}, \label{eq_safty_high_order} \\
|x_i(t)| \le \max\{\breve{\beta}_i(&\max\{\breve{V}(x_1(0)), |\rho_0(x_1(0))|, \notag \\
&|x_2(0)|,\ldots,|x_m(0)|\},t),~ \breve{c}_i\} \label{eq_xi_boundness}
\end{align}
\end{subequations}
for $t\in\mathcal{I}$ and $i=2,\ldots,m$, where $\breve{V}(x_1)=\max_{j=1,\ldots,n_c}V_j(x_1)$, and $\mathcal{I}\subseteq\mathbb{R}_+$ is the time interval over which $z_m(t)$ is right maximally defined\footnote{Here, the state trajectory is defined as the solution to the initial-value problem in the sense of Carath{\'{e}}odory \cite{Cortes-CSM-2008, Glotfelter-Cortes-Egerstedt-TAC-2020}, and the uniqueness is not demanded.}. 

\begin{figure}[!ht]
\centering
\begin{tikzpicture}[scale=1, transform shape]
\draw[rounded corners=4pt, dashed, fill=green!5!white] (-9.1,1.5) rectangle (-0.1,-1);

% Rectangle for Plant
\draw[rounded corners=4pt, fill=gray!20!white, line width=0.75pt] (0.4,0.7) rectangle (1.8,-0.7);
% Node and equation label for Plant
\node at (1.1,0.2) {Plant};
\node at (1.1,-0.2) {\eqref{eq_plant_dynamics}};

% Rectangle for Safety Controller 1
\draw[rounded corners=4pt, fill=green!20!white, line width=0.75pt] (-8.5,0.7) rectangle (-7.1,-0.7);
% \node at (-8.7,-0.2) {\eqref{eq_safety_control_law_varrhoL}};

% Rectangle for Safety Controller 2
\draw[rounded corners=4pt, fill=green!20!white, line width=0.75pt] (-6,0.7) rectangle (-4.6,-0.7);
% \node at (-5.3,-0.2) {\eqref{eq_safety_control_law_varrhoL}};

% Rectangle for Safety Controller 3
\draw[rounded corners=4pt, fill=green!20!white, line width=0.75pt] (-2.5,0.7) rectangle (-1.1,-0.7);
% \node at (-1.6,-0.2) {\eqref{eq_safety_control_law_varrhoL}};

% Arrows and labels
\draw[-latex] (-1.1,0) -- (0.4,0);
\node at (0.1,0.3) {$u$};

\draw[-latex] (1.8,0.5) -- (2.5,0.5) -- (2.5,-2) -- (-9.4,-2) -- (-9.4,-0.3) -- (-8.5,-0.3);
\node at (-9.6,-1.7) {$x_1$};

% Ellipse for x_2
\draw (-6.4,0) ellipse (0.1 and 0.1);
\draw[-latex] (-7.1,0) -- (-6.5,0);

% Arrows for x_2, x_3, ..., x_n
\draw[-latex] (1.8,0.3) -- (2.3,0.3) -- (2.3,-1.7) -- (-6.4,-1.7) -- (-6.4,-0.1);
\draw[-latex] (-6.3,0) -- (-6,0);
\draw[-latex] (-4.6,0) -- (-4.1,0);
\draw[-latex] (-2.9,0) -- (-2.5,0);
\draw (-3,0) ellipse (0.1 and 0.1);
\draw[-latex] (1.8,-0.4) -- (2.1,-0.4) -- (2.1,-1.4) -- (-3,-1.4) -- (-3,-0.1);
\draw[-latex] (-3.6,0) -- (-3.1,0);

% Additional labels and ellipses
\node at (-3.8,0) {$\cdots$};
\node at (-6.7,-1.5) {$x_2$};
\node at (-3.4,-1.2) {$x_m$};
\node at (-7.8,0) {$\rho_1$};
\node at (-5.3,0) {$\rho_2$};
\node at (-1.8,0) {$\rho_m$};

% Nodes for x^* and \tilde{x}
\node at (-6.8,0.3) {$x^*_2$};
\node at (-4.3,0.3) {$x^*_3$};
\node at (-3.4,0.3) {$x^*_m$};

% Operator Symbols
\node at (-6.6,-0.4) {$+$};
\node at (-3.2,-0.4) {$+$};
\node at (-6.7,-0.2) {$-$};
\node at (-3.4,-0.2) {$-$};
\node at (2,0.1) {$\vdots$};
\node[align=left] at (-7.2,1.1) {Safety Control Law \eqref{eq_controller_high}};

\draw[rounded corners=4pt, dashed, fill=cyan!5!white] (-6.9,3.6) rectangle (-0.1,1.7);
\draw[rounded corners=4pt, fill=cyan!20!white, line width=0.75pt] (-2.5,3.4) rectangle (-1.1,2);
\node[align=left] at (-4.8,3.1) {Nominal Controller \eqref{eq_def_nominal_controller_highorder}};
\node at (-1.8,2.7) {$\rho_0$};
\draw[-latex] (2.5,0.5) -- (2.5,2.7) -- (-1.1,2.7);
\draw[-latex] (-2.5,2.7) -- (-9.4,2.7) -- (-9.4,0.3) -- (-8.5,0.3);

\draw  (2.5,0.5)[fill=black] ellipse (0.03 and 0.03);
\node at (-2.9,2.4) {$u_0$};
\node at (-0.6,0.3) {$x^*_{m+1}$};
\end{tikzpicture}
\caption{Block diagram of the proposed safety control system.}
\label{fig_block_diagram_high_relative}
\end{figure}
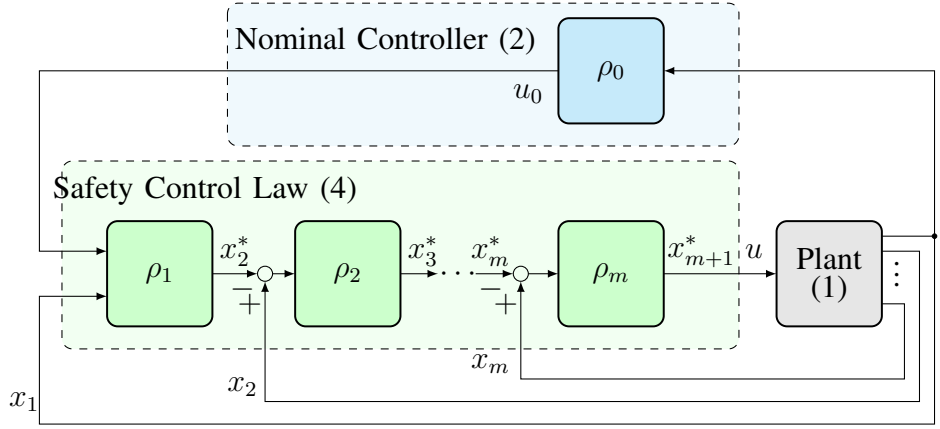

The block diagram of the proposed safety control system is shown in Figure \ref{fig_block_diagram_high_relative}.

We make the following assumptions on the plant and the certificate functions.
\begin{assumption}\label{assumption_high_relative_degree}
For the plant \eqref{eq_plant_dynamics}, there exist nonnegative constants $\ubar{g}$, $\bar{g}$ and $\bar{\delta}$ and functions $\bar{f}_{x_1},\ldots,\bar{f}_{x_m}\in\mathcal{K}\cup\{0\}$ such that
\begin{subequations}\label{assume_system_model_highorder}
\begin{align}
|f_i(z_i)| &\le \sum^i_{j=1}\bar{f}_{x_j}(|x_j|),\label{eq_assume_f_max_highorder}\\
\!\!\lambda_{\min}^{\frac{1}{2}}(g_ig^T_i) &\ge \ubar{g}, \label{eq_assume_g_min_highorder} \\
|g_i|&\le \bar{g}, \label{eq_assume_g_max_highorder}\\
|\delta_i(z_i)| &\le \bar{\delta},\label{eq_assume_delta_max_highorder}
\end{align}
\end{subequations}
for all $z_m\in\mathbb{R}^{n_{z_m}}$.
\end{assumption}

\begin{assumption}\label{assumption_certificate_functions}
For $j=1,\ldots,n_c$, there exist positive constants $v_j$ such that
\begin{align}
\{x\in\mathcal{X}:V_j(x)\ge v_j \text{~and~} V_k(x)\ge v_k\} = \emptyset, \label{eq_assume_unsafety_set_nointersection}
\end{align}
for all $k\neq j$. Moreover, for each $j=1,\ldots,n_c$, the certificate function $V_j:\mathcal{X}\rightarrow\mathbb{R}_+$ is continuously differentiable on $\mathcal{X}$, and its gradient satisfies 
\begin{align}
\frac{\partial V_j(x)}{\partial x} &\neq 0, ~~~\forall x\in\mathcal{X}, \label{eq_assume_nozero_gradient}\\
\left|\frac{\partial V_j(x)}{\partial x}\right| &\ge \eta, ~~~\text{wherever~} V_j(x)\ge v_j \text{~and~} x\in\mathcal{X}
\end{align}
with $\eta$ being a positive constant.
\end{assumption}

\begin{remark}
With $\{x\in\mathcal{X}:V_j(x)\ge v_j\}$ considered as the unsafe set characterized by $V_j$ and $v_j$, condition \eqref{eq_assume_unsafety_set_nointersection} means that different unsafe sets are disjoint. From the view point of robustness analysis, such assumption guarantees the existence of available gradients for safety control in the presence of perturbations.
\end{remark}

\begin{assumption}\label{assumption_nominal_controller}
The nominal controller $\rho_0:\mathbb{R}^{n_x}\rightarrow\mathbb{R}^{n_u}$ is continuously differentiable, and there exists a constant $\bar{\rho}_0>0$ such that
\begin{align}
|\rho_0(x)|&\le\bar{\rho}_0\label{eq_assume_rho_r_upperbound}
\end{align}
for $x\in\mathbb{R}^{n_x}$.
\end{assumption}

\begin{remark}
Although the nominal controllers, represented by $\rho_0$ in this paper, does not explicitly depend on time, our main result can be easily extended to more general cases in which time is also an argument of $\rho_0$, as long as mild boundness and Lipschitz continuity conditions are satisfied.
\end{remark}

\begin{remark}
For the class of nonlinear uncertain cascade systems subject to multiple constraints, existing approaches that integrate control barrier functions with optimization face challenges. These challenges arise from the complexities of accurately determining gradients corresponding to control input and selecting parameters across multiple hierarchical levels, making it difficult to simultaneously guarantee the safety of the plants and the feasibility of the optimization.
\end{remark}

If the safety certificate functions are reduced to a candidate Lyapunov function, then the safety control problem is reduced to the standard stabilization problem. 
Backstepping techniques can be readily applied and the resulting controller naturally features the structure of \eqref{eq_controller_high} \cite{Jiang-Automatica-1999, Sepulchre-Jankovic-Kokotovic-Book-1997, Jiang-Mareels-TAC-1997}.

\subsection{A Sub-Problem: Systems with Relative-Degree-One}
\label{section_problem_formulation_relative_degree_one}

% The first step in solving the safety control problem for nonlinear plants in cascade form is to address the safety control of nonlinear plants with a relative degree of one. 
When $m=1$, the plant \eqref{eq_plant_dynamics} is reduced to the following form:
\begin{align}
\dot{x}=f(z,x)+(g+\delta(z,x))(u+w),\label{eq_model_relative_degree_one}
\end{align}
where $x\in\mathbb{R}^{n_x}$ is the state, $u\in\mathbb{R}^{n_u}$ is the control input, $z\in\mathbb{R}^{n_z}$ and $w\in\mathbb{R}^{n_u}$ are external disturbance inputs injected into the plant through different channels, $g\in\mathbb{R}^{n_x \times n_u}$ is a constant matrix.
The functions $f:\mathbb{R}^{n_z}\times\mathbb{R}^{n_x}\rightarrow\mathbb{R}^{n_x}$ and $\delta:\mathbb{R}^{n_z}\times\mathbb{R}^{n_x}\rightarrow\mathbb{R}^{n_x\times n_u}$ are locally Lipschitz. When considered as functions of time $t$, the disturbance inputs $z$ and $w$ are assumed to be piecewise continuous. The subscripts are omitted to simplify the notations.

Given a nominal controller 
\begin{align}
	u_0=\rho_0(x) \label{eq_nominal_controller}
\end{align}
with $u_0\in\mathbb{R}^{n_u}$ representing the nominal control input and $\rho_0:\mathbb{R}^{n_x}\rightarrow\mathbb{R}^{n_u}$, and safety certificate functions $V_j:\mathcal{X}\rightarrow\mathbb{R}_+$ for $j=1,\ldots,n_c$ with $\mathcal{X}\subseteq\mathbb{R}^{n_x}$ being a nonempty open set, we expect to design a continuous safety controller in the form of
\begin{align}
	u=\rho(x)\label{eq_controller_relative_degree_one}
\end{align}
with $\rho:\mathbb{R}^{n_x}\rightarrow\mathbb{R}^{n_u}$ depending on $\rho_0$, such that the actual control input $u$ minimally invades the nominal control input $u_0$, and the closed-loop system \eqref{eq_model_relative_degree_one} and \eqref{eq_controller_relative_degree_one} is ISpSf in the sense that the state trajectory $x(t)$ keeps inside $\mathcal{X}$ whenever it is defined, and there exist $\beta\in\mathcal{KL}$, $\gamma^w,\gamma^z\in\mathcal{K}$ and constant $c_j\ge 0$ such that for any initial state $x(0)\in\mathcal{X}$ and any piecewise continuous $z$ and $w$,
\begin{align}
	V_j(x(t)) \le\max\{\beta(V_j(x(0)), t), \gamma^w(\|w\|_t), \gamma^z(\|z\|_t), c_j\}, \label{eq_objective_ISpS_safety}
\end{align}
for all $t\in\mathcal{I}$ and $j=1,\ldots,n_c$, where $\mathcal{I}\subseteq\mathbb{R}_+$ is the time interval over which $x(t)$ is right maximally defined. 

In accordance with Assumption \ref{assumption_high_relative_degree}, the following assumption is made on the plant \eqref{eq_model_relative_degree_one}.

\begin{assumption}\label{assumption_model_relative_degree_one}
For the plant \eqref{eq_model_relative_degree_one}, there exist nonnegative constants $\ubar{g}$, $\bar{g}$ and $\bar{\delta}$ and functions $\bar{f}_z,\bar{f}_x\in\mathcal{K}\cup\{0\}$ such that 
\begin{subequations} \label{assume_system_model}
\begin{align}
|f(z,x)| &\le \bar{f}_z(|z|)+\bar{f}_x(|x|), \label{eq_assume_f_max}\\
\lambda_{\min}^{\frac{1}{2}}(gg^T) &\ge \ubar{g}, \label{eq_assume_g_min} \\
|g|&\le \bar{g}, \label{eq_assme_g_max}\\
|\delta(z,x)| &\le \bar{\delta},\label{eq_assume_delta_max}
\end{align}
\end{subequations}
for $z\in\mathbb{R}^{n_z}$ and $x\in\mathbb{R}^{n_x}$.
\end{assumption}

\section{Design Ingredient: Safety Control of Nonlinear Plants with Relative-Degree-One}
\label{section_relative_degree_one}

This section studies the safety control problem for a class of nonlinear uncertain systems with relative-degree-one (see Section \ref{section_problem_formulation_relative_degree_one}). The problem setting in this section takes into account multiple safety constraints and external disturbances. In particular, each safety constraint is characterized by a state-dependent safety certificate function. By safety, we mean that the value of each safety certificate function decays with respect to time whenever larger than a desired threshold value depending on the magnitudes of the disturbances.

For a plant in the control-affine form, the gradients of the safety certificate functions naturally result in linear constraints on the control input, and one may select a control input by solving a QP problem in real time \cite{Ames-Coogan-Egerstedt-ECC-2019}. 
For plants subject to uncertainties and disturbances, QCQP-based algorithms \cite{Castaneda-Choi-Zhang-Tomlin-Sreenath-ACC-2021, Cosner-Singletary-Taylor-Molnar-Bouman-Ames-IROS-2021} naturally extend the QP-based scheme. 
In this section, we consider a safety-control-oriented QCQP problem that incorporates multiple safety constraints and uncertainties and prove its feasibility when the plant dynamics and the safety constraints satisfy mild conditions. The tool presented in this section plays an instrumental role for constructive safety control of nonlinear plants of higher relative degrees, see Section \ref{section_main_result}.

\subsection{Safety Controller Design}
\label{subsection_relative_degree_one_robust_safety_control}

Given the certificate functions $V_j$ for $j=1,\ldots,n_c$, define
\begin{align}
&\mathcal{F}(z,x,w) := \Big\{u\in\mathbb{R}^{n_u}: \frac{\partial V_j(x)}{\partial x}\Big((g+\delta(z,x))(u+w)+\notag\\
&~f(z,x)\Big)\le -\alpha_j(V_j(x)-v_j), j=1,\ldots,n_c\Big\}, \label{eq_ideal_control_input_set}
\end{align}
for $z\in\mathbb{R}^{n_z}$, $x\in\mathcal{X}$ and $w\in\mathbb{R}^{n_u}$. It can be observed that the set $\mathcal{F}(z,x,w)$ is defined by linear constraints characterized by the gradient functions of the certificate functions $V_j$. 

For the plant \eqref{eq_model_relative_degree_one}, if one chooses $\alpha_j\in\mathcal{K}^e$ for $j=1,\ldots,n_c$ such that the image of the set-valued map $\mathcal{F}$ is nonempty wherever it is defined, and set
\begin{align}
u\in\mathcal{F}(z,x,w), \label{eq_set_valued_control_ideal}
\end{align}
then there exist $\beta\in\mathcal{KL}$, $\gamma^w,\gamma^z\in\mathcal{K}$ and $c_j>v_j$ for the achievement of the control objective \eqref{eq_objective_ISpS_safety}. See \cite{Aubin-Frankowska-1990-book} for notations of set-valued analysis.

However, condition \eqref{eq_set_valued_control_ideal} cannot be readily verified due to the unknown functions $f$ and $\delta$, as well as the unknown disturbances $z$ and $w$. Taking into account the boundedness properties of the unknown terms (see Assumption \ref{assumption_model_relative_degree_one}), one would solve the problem in a robust way. In particular, we define
\begin{align}
\mathcal{F}_c(x):=\left\{u\in\mathbb{R}^{n_u}:A_c(x)u+\frac{\bar{\delta}}{\ubar{g}}|u|\le b_c(x)\right\},\label{eq_practical_control_input_set}
\end{align}
where 
\begin{subequations}\label{eq_practical_control_input_set_Acbc}
\begin{align}
[A_c(x)]_{j,:} &= \frac{\partial V_j(x)}{\partial x} g\left|\frac{\partial V_j(x)}{\partial x} g\right|^{-1}, \label{eq_practical_control_input_set_A_c}\\
[b_c(x)]_j &= -\alpha_j(V_j(x)-v_j), \label{eq_practical_control_input_set_b_c}
\end{align}
\end{subequations}
with each row of $A_c(x)$ being normalized for the convenience of expressions and proofs.
Then, the expected property
\begin{align}
\nabla V_j(x) \dot{x} &\le -\left|\frac{\partial V_j(x)}{\partial x}g\right|\Big(\alpha_j(V_j(x)-v_j) \notag \\
&\left.~~~-\frac{1}{\ubar{g}}\bar{f}_z(|z|) -\frac{1}{\ubar{g}}\bar{f}_x(|x|) - \left(1+\frac{\bar{\delta}}{\ubar{g}}\right)|w|\right) \label{eq_practical_dissipation}
\end{align}
holds for all $j=1,\ldots,n_c$, if we choose $\alpha_j\in\mathcal{K}^e$ for $j=1,\ldots,n_c$ such that the image of the set-valued map $\mathcal{F}_c$ is nonempty wherever it is defined, and set
\begin{align}
u\in\mathcal{F}_c(x) \label{eq_set_valued_control_practical}
\end{align}
(see Appendix \ref{appendix_proof_smallerfeasibleset} for a sketch of the proof).

Here, we would like to mention that the employment of the new set-valued map $\mathcal{F}_c$ defined by \eqref{eq_practical_control_input_set} to address the uncertainties naturally results in quadratic constraints.

Incorporating the nominal controller $\rho_0$, we formulate the safety control problem as the following QCQP problem:
\begin{subequations}\label{eq_safety_control_law}
\begin{align}
&u=\argmin_u|u-\rho_0(x)|^2, \label{eq_safety_control_law1}\\
&\text{s.t.}~u\in\mathcal{F}_c(x).\label{eq_safety_control_law2}
\end{align}
\end{subequations}

\begin{figure}[!ht]
\centering
\begin{tikzpicture}[scale=1, transform shape]
% Rectangle for Plant
\draw[rounded corners=4pt, fill=gray!20!white, line width =0.75pt]  (1,0.4) rectangle (3.4,-0.7);
% Node and equation label for Plant
\node[align=center] at (2.2,-0.2) {Plant \eqref{eq_model_relative_degree_one}};

% Rectangle for Safety Controller
\draw[rounded corners=4pt, fill=green!20!white, line width =0.75pt]  (-4.2,0.4) rectangle (-0.2,-0.7);
% Node and equation label for Safety Controller
\node[align=center] at (-2.2,-0.2) {Safety Controller \eqref{eq_safety_control_law}};

% Rectangle for Nominal Controller
\draw[rounded corners=4pt, fill=cyan!20!white, line width =0.75pt]  (-2.7,1.9) rectangle (2.1,0.9);
\node[align=center] at (-0.2,1.4) {Nominal Controller \eqref{eq_nominal_controller} };

% Arrows and labels
\draw[-latex] (-0.2,-0.2) -- (1,-0.2);
\node at (0,0) {$u$};

\node at (0.4,-0.7) {$w$};
\node at (0.4,0.3) {$z$};

\draw[-latex] (3.4,-0.2) -- (4,-0.2)-- (4,-1.1) -- (-4.6,-1.1)  -- (-4.6,-0.5) -- (-4.2,-0.5);
\draw[-latex] (0.4,-0.5) -- (1,-0.5);
\node at (3.7,0) {$x$};

\draw [fill=black] (4,-0.2) ellipse (0.05 and 0.05);

\draw[-latex] (-2.7,1.4) --(-4.6,1.4) --(-4.6,0) -- (-4.2,0);

\draw[-latex]  (0.4,0.1) -- (1,0.1);
\draw (4,-0.2) -- (4,1.4) -- (2.1,1.4);
\node at (-3.3,1.7) {$u_0$};
\end{tikzpicture}
\caption{Block diagram of the safety control system.}
\label{figure_block_diagram_relative_one}
\end{figure}
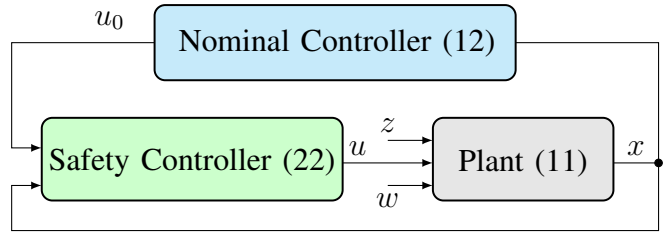

Figure \ref{figure_block_diagram_relative_one} shows the resulting safety control system.

% It is expected that the QCQP problem \eqref{eq_safety_control_law} results in a feasible safety controller in the form of
% \begin{align}
% u=\rho(x).
% \end{align}

Before presenting the main result of this section, we give some properties of the safety certificate functions satisfying Assumption \ref{assumption_certificate_functions}.

\begin{lemma} \label{lemma_certificate_functions}
Consider the certificate function $V_j$ and the corresponding positive constant $v_j$ for $j=1,\ldots,n_c$. The satisfaction of condition \eqref{eq_assume_nozero_gradient} implies the existence of $\ubar{\alpha}_j,\bar{\alpha}_j\in\mathcal{K}^e$ defined on $(-a,b)$ such that
\begin{align}
&\{D_j(x):x\in\mathcal{X}\}\subseteq (-a,b), \label{eq_assume_certificate_range} \\
&\lim_{s\rightarrow b}\bar{\alpha}_j(s)\rightarrow \infty, \label{eq_assume_certificate_limits}\\
&\ubar{\alpha}_j(D_j(x))\le V_j(x)-v_j \le \bar{\alpha}_j(D_j(x)), ~~~\forall~x\in\mathcal{X}, \label{eq_assume_certificate_bound}
\end{align}
where
\begin{align}
D_j(x) = \sgn(V_j(x)-v_j) \inf\limits_{s\in\{x:V_j(x)= v_j\}}|x-s|,
\label{eq_assume_dist_barrierfunction}
\end{align}
for $j=1,\ldots,n_c$.
\end{lemma}

See Appendix \ref{appendix_proof_certificate_functions} for the proof.

Theorem \ref{theorem_safety_relative_degree_one} shows the validity of the QCQP-based algorithm to solve the safety control problem.

\begin{theorem}\label{theorem_safety_relative_degree_one}
Consider the plant \eqref{eq_model_relative_degree_one}, the certificate functions $V_1,\ldots, V_{n_c}$ and the QCQP-based controller \eqref{eq_safety_control_law} with $A_c$, $b_c$ and $\alpha_1,\ldots,\alpha_{n_c}$ defined in \eqref{eq_practical_control_input_set_Acbc}. Suppose that Assumptions \ref{assumption_certificate_functions}, \ref{assumption_nominal_controller} and \ref{assumption_model_relative_degree_one} are satisfied, and moreover 
\begin{align}
\ubar{g}>\bar{\delta}. \label{eq_ubarg_bardelta}
\end{align}
For $j=1,\ldots,n_c$, choose
\begin{align}
\alpha_j(s)=\alpha\circ\bar{\alpha}^{-1}_j(s) \label{eq_alpha_j_definition}
\end{align}
for $s\in (-v_j,\infty)$, where $\alpha\in\mathcal{K}^e$ satisfies
\begin{align}
\alpha(s) + \frac{\ubar{g}+\bar{\delta}}{\ubar{g}-\bar{\delta}}\alpha(-s)\le 0 \label{eq_feasibility_condition}
\end{align}
for all $s\le 0$ and the functions $\ubar{\alpha}_j, \bar{\alpha}_j\in\mathcal{K}^e$ are given by Lemma \ref{lemma_certificate_functions}. Then, it holds that
\begin{enumerate}
\item $\mathcal{F}_c(x)$ defined by \eqref{eq_practical_control_input_set} is non-empty for each $x\in\mathcal{X}$;
\item 
for $j=1,\ldots,n_c$, given constants $c_j>v_j$, $\theta>0$ and functions $\gamma^{w},\gamma^{z}\in\mathcal{K}$, if the function $\alpha_j$ is chosen to further satisfy
\begin{align}
&\alpha_j\Big(\frac{c_j-v_j}{c_j}V_j(x)\Big)
% [b_c(x)]_j 
\ge \theta V_j(x) + \frac{1}{\ubar{g}}\bar{f}_z\circ(\gamma^z)^{-1}(V_j(x))  \notag \\
&~~~+ (1+\frac{\bar{\delta}}{\ubar{g}})(\gamma^w)^{-1}(V_j(x)) + \frac{1}{\ubar{g}}\bar{f}_x(|x|) \label{eq_chosen_of_bc}
\end{align}
for all $x\in\{x:V_j(x)\ge c_j\}$, then the certificate function $V_j:\mathcal{X}\rightarrow\mathbb{R}_+$ satisfies 
\begin{align}
V_j(x(t)) \le \max\{&\beta(V_j(x(0)), t), \gamma^w(\|w\|_t),\gamma^z(\|z\|_t), c_j\} \label{eq_main_conclusion_ISpS}
\end{align}
for all $t\in\mathcal{I}$, with $\beta(s,t)=s e^{-\eta\ubar{g}\theta t}$.
\end{enumerate}
\end{theorem}

The proof of Theorem \ref{theorem_safety_relative_degree_one} is given in Section \ref{subsection_proofs_of_safety_controller}. 

\begin{remark}\label{remark_theorem_conditions_satisfactions}
One can always find a $\alpha_j$ to satisfy the conditions required by Theorem \ref{theorem_safety_relative_degree_one} as long as \eqref{eq_chosen_of_bc} can be satisfied by some $\alpha_j$. If $\bar{f}_x$ is bounded, condition \eqref{eq_chosen_of_bc} can be trivially satisfied.
\end{remark}

\begin{remark}\label{remark_ISpSf_and_ZCBF}
The employment of safety certificate functions would make it more convenient to characterize the safety property of the closed-loop system subject to disturbance inputs. Barrier functions in other forms used in the existing literature \cite{Ames-Coogan-Egerstedt-ECC-2019, Ames-Xu-Grizzle-Tabuada-TAC-2017} can be transformed into our form. As an example, for plant $\dot{x}=u$, given 
\begin{align}
\mathcal{C}   &= \{x\in\mathcal{X}:h(x)\ge 0\}, \\
\mathcal{U}(x) &= \Big\{u\in\mathbb{R}^{n_x}:\frac{\partial h(x)}{\partial x}u \ge -\alpha(h(x))\Big\}, \label{eq_cbf_feasibleset}
\end{align}
which are characterized by a zeroing control-barrier function $h:\mathcal{X}\rightarrow\mathbb{R}$ with $\alpha\in\mathcal{K}^e$. Define 
\begin{align}
V(x) = \mu(h(x)),
\end{align}
where $\mu$ is a continuously differentiable, strictly decreasing, strictly convex function satisfying  $\lim_{s\rightarrow\infty}\mu(s)=0$. Then, we can equivalently represent $\mathcal{C}$ and $\mathcal{U}(x)$ with $V$ as 
\begin{align}
\mathcal{C} &= \{x\in\mathcal{X}:V(x)\le \mu(0)\}, \\
\mathcal{U}(x) &= \Big\{u\in\mathbb{R}^{n_x}:\frac{\partial V(x)}{\partial x}u \le -\alpha'(V(x)-\mu(0))\Big\}, 
\end{align}
where $\alpha'(s) =-\alpha_{\mu}(s+\mu(0))\alpha(\mu^{-1}(s+\mu(0)))$ with $\alpha_{\mu}$ defined as:
\begin{align}
\alpha_{\mu}(s) =\begin{cases}
-\frac{\mathrm{d} \mu(\mu^{-1}(s))}{\mathrm{d} \mu^{-1}(s)}, &\text{if } s>0,\\
0, &\text{otherwise}.
\end{cases}
\end{align}
Notice that the function $\alpha'$ is of class $\mathcal{K}^e$.
\end{remark}

%\begin{remark}
%Several methods, such as those discussed in \cite{Cosner-Singletary-Taylor-Molnar-Bouman-Ames-IROS-2021, Castaneda-Choi-Zhang-Tomlin-Sreenath-ACC-2021}, have been developed for safety control of nonlinear plants in the form of \eqref{eq_model_relative_degree_one}. 
%These results take into account robustness issues with respect to uncertain dynamics and external disturbances, but with constraints limited to no more than two. 
%Also, to the best of our knowledge, there seem rare constructive results based on QCQP algorithms with ensured feasibility in the presence of uncertainties and multiple constraints.
%\end{remark}

\subsection{Proof of Theorem \ref{theorem_safety_relative_degree_one}}
\label{subsection_proofs_of_safety_controller}

\subsubsection{Property 1)}\label{subsubsection_proof_of_robust_safety}

For each $x\in\mathcal{X}$, we show that $\mathcal{F}_c(x)$ defined by \eqref{eq_practical_control_input_set} is nonempty. For this purpose, we consider the following two cases:
\begin{itemize}
\item Case 1: $\min_{j=1,\ldots,n_c}[b_c(x)]_{j}\ge 0$. In this case, $0$ is always an element of $\mathcal{F}_c(x)$.
\item Case 2: $\min_{j=1,\ldots,n_c}[b_c(x)]_{j}< 0$. In this case, we set
\begin{align}
u_*(x)= [A_c(x)]_{\ubar{j},:}^T\dfrac{[b_c(x)]_{\ubar{j}}}{1-\bar{\delta}/\ubar{g}}, \label{eq_proof_u_star}
\end{align}
where $\ubar{j}$ represents the index of any one of the smallest elements of $b_c(x)$, that is, $[b_c(x)]_{\ubar{j}}=\min_{j=1,\ldots,n_c}[b_c(x)]_{j}$, and prove $u_*(x)\in\mathcal{F}_c(x)$ by showing the simultaneous satisfaction of all the constraints in $\mathcal{F}_c(x)$:
\begin{itemize}
\item $u_*(x)$ satisfies the $\ubar{j}$-th constraint in $\mathcal{F}_c(x)$. Substituting $u_*(x)$ into the $\ubar{j}$-th inequality of $\mathcal{F}_c(x)$, we have 
\begin{align}
[A_c(x)]_{\ubar{j},:}u_*(x)+ \frac{\bar{\delta}}{\ubar{g}}|u_*(x)|
&=\Big(1-\frac{\bar{\delta}}{\ubar{g}}\Big)\frac{[b_c(x)]_{\ubar{j}}}{1-\bar{\delta}/\ubar{g}} \notag \\
&= [b_c(x)]_{\ubar{j}}. \label{eq_feasibility_proof_relationship1}
\end{align}

\item $u_*(x)$ satisfies other constraints in $\mathcal{F}_c(x)$. From \eqref{eq_assume_unsafety_set_nointersection}, applying the triangle inequality \cite[Theorem 3.3]{Apostol-book-1973} yields:
\begin{align}
\!\!\max\{D_j(x), D_{\ubar{j}}(x)\} + \min\{D_j(x), D_{\ubar{j}}(x)\}\le 0 \label{eq_certificate_distance}
\end{align}
holds for all $x\in\mathcal{X}$, where $D_j:\mathcal{X}\rightarrow\mathbb{R}$ is defined in \eqref{eq_assume_dist_barrierfunction}. 
For any $x\in\mathcal{X}$, define
\begin{align}
\Delta(x) &= - \frac{\ubar{g}+\bar{\delta}}{\ubar{g}-\bar{\delta}}[b_c(x)]_{\ubar{j}}-[b_c(x)]_j.\label{eq_def_delta_x}
\end{align}
Then, direct calculation leads to
\begin{align}
\Delta(x)&\le \frac{\ubar{g}+\bar{\delta}}{\ubar{g}-\bar{\delta}} \alpha(D_{\ubar{j}}(x)) +\alpha(D_j(x))\notag \\
&\le \frac{\ubar{g}+\bar{\delta}}{\ubar{g}-\bar{\delta}} \alpha(\max\{D_j(x),D_{\ubar{j}}(x)\}) \notag \\
&~~~+\alpha(\min\{D_j(x),D_{\ubar{j}}(x)\}) \notag \\
&\le \frac{\ubar{g}+\bar{\delta}}{\ubar{g}-\bar{\delta}} \alpha(-\min\{D_j(x),D_{\ubar{j}}(x)\})\notag \\
&~~~+ \alpha(\min\{D_j(x),D_{\ubar{j}}(x)\}) \le 0, \label{eq_proof_delta_x}
\end{align}
where we have used the definition of $b_c(x)$ in \eqref{eq_practical_control_input_set_b_c} and inequality \eqref{eq_assume_certificate_bound} for the first inequality, used \eqref{eq_certificate_distance} for the third inequality, and used \eqref{eq_feasibility_condition} for the last inequality.

Then, substituting $u_*(x)$ into the $j$-th constraint in $\mathcal{F}_c(x)$ with $j\neq \ubar{j}$, we have 
\begin{align}
& [A_c(x)]_{j,:}u_*(x) + \frac{\bar{\delta}}{\ubar{g}}|u_*(x)| \le (1+\frac{\bar{\delta}}{\ubar{g}})|u_*(x)| \notag \\
&=-\frac{\ubar{g}+\bar{\delta}}{\ubar{g}-\bar{\delta}} [b_c(x)]_{\ubar{j}}
= [b_c(x)]_j + \Delta(x) \notag \\
&\le  [b_c(x)]_j, \label{eq_feasibility_proof_relationship2}
\end{align}
where we have used the triangle inequality again for the first inequality, used the definition of $u_*$ in \eqref{eq_proof_u_star} for the first equality, used the definition of $\Delta(x)$ in \eqref{eq_def_delta_x} for the second equality, and used \eqref{eq_proof_delta_x} for the last inequality.
\end{itemize}
\end{itemize}
Based on properties \eqref{eq_feasibility_proof_relationship1} and \eqref{eq_feasibility_proof_relationship2}, we conclude that for each $x\in\mathcal{X}$, the set $\mathcal{F}_c(x)$ is non-empty.

\subsubsection{Property 2)}
For any $u\in \mathcal{F}_c(x)$, if $V_j(x)\ge c_j$, we have
\begin{align}
V_j(x) - v_j \ge \frac{c_j-v_j}{c_j} V_j(x). \label{eq_proof_ISpS_01}
\end{align}

Then, taking the derivative of $V_j$ along the trajectories of the closed-loop system composed of \eqref{eq_model_relative_degree_one} and \eqref{eq_safety_control_law}, we have
\begin{align}
\nabla V_j \dot{x} &\le-\left|\frac{\partial V_j(x)}{\partial x}g\right|\Big(-\alpha_j(V_j(x)-v_j) -\frac{1}{\ubar{g}}\bar{f}_x(|x|) \notag \\
&~~~-\frac{1}{\ubar{g}}\bar{f}_z(|z|) - \big(1+\frac{\bar{\delta}}{\ubar{g}}\big)|w|)\Big) \notag \\
&\le -\left|\frac{\partial V_j(x)}{\partial x}g\right|(\theta V_j(x) + \bar{f}_z\circ(\gamma^z)^{-1}(V_j(x))  \notag \\
&~~~- \bar{f}_z(|z|) + \big(1+\frac{\bar{\delta}}{\ubar{g}}\big)((\gamma^w)^{-1}(V_j(x)) - |w|)),
\end{align}
where we have used \eqref{eq_practical_dissipation} for the first inequality, and used \eqref{eq_chosen_of_bc} and \eqref{eq_proof_ISpS_01} for the second inequality.
It can be verified that 
\begin{align}
V_j(x) &\ge \max\{\gamma^w(|w|),\gamma^z(|z|), c_j\}, \notag \\
\Rightarrow \nabla V_j \dot{x} &\le -\left|\frac{\partial V_j(x)}{\partial x}g\right|\theta V_j(x) \le -\eta\ubar{g}\theta V_j(x).
\end{align}

Based on the discussion above, using the comparison principle \cite{Khalil-book-2002}, we can prove the satisfaction of property \eqref{eq_main_conclusion_ISpS} with $\beta(s,t)= s e^{-\eta\ubar{g}\theta t}$ \cite{Sontag-2008}. 
This ends the proof of Theorem \ref{theorem_safety_relative_degree_one}.

\section{Design Ingredient: Reshaping Feasible Set of QCQP}
\label{section_feasible_set_reshaping}

The Lipschitz continuity issues of the solutions to optimization problems have attracted considerable attention from the literature of numerical optimization \cite{Morris-Powell-Ames-CDC-2015, Ong-Cortes-CDC-2019, Chan-Mar-SIAM-2017}. 
For QP-based safety control, Lispchitz continuity of the controller is highly demanded for practical implementations with enhanced robustness \cite{Ames-Coogan-Egerstedt-ECC-2019, Ames-Xu-Grizzle-Tabuada-TAC-2017, Morris-Powell-Ames-CDC-2015, Wu-Liu-Egerstedt-Jiang-2023-TAC}. 
In our case, the Lipschitz continuity property is crucial in the expected constructive safety control design procedure for guaranteed robustness with respect to uncertain nonlinear dynamics associated with different relative degrees. 

In this section, we study the more general QCQP problem arising from safety control in the presence of uncertain dynamics and disturbances as discussed in Section \ref{section_relative_degree_one}. Our major contribution lies in the refined feasible-set reshaping technique. The basic idea is to project the quadratic constraints involving uncertainties onto a predefined positive basis, which results in linear constraints and a QP problem. Under mild conditions, by appropriately choosing the positive basis, we can guarantee that the nonredundant active constraint matrix is never singular and prove that the solution to the resulting QP problem is Lipschitz with a predetermined Lipschitz constant.

\subsection{The Non-Lipschitz Issue with Normal QCQP}
\label{subsection_feasible_set_reshaping_nonlipshcitz_issue}
We consider the following QCQP problem arising from safety control:
\begin{subequations}\label{eq_nonLipschitz_Optimization}
\begin{align}
&u = \argmin_u |u-\rho_0(x)|^2 \label{eq.Lipschitz.Optimization}\\
&\text{s.t.}~u \in\mathcal{U}_c(x),
\end{align}
\end{subequations}
where $x\in\mathcal{X}\subseteq\mathbb{R}^{n_x}$ is the system state, $\rho_0:\mathbb{R}^{n_x}\rightarrow \mathbb{R}^{n_u}$ represents the nominal controller, and $\mathcal{U}_c:\mathbb{R}^{n_x}\leadsto\mathbb{R}^{n_u}$ represents the original feasible set, which is defined by
\begin{align}
\mathcal{U}_c(x) &= \{u\in\mathbb{R}^{n_u}:A_c(x)u + c_c(x)|u| \le b_c(x)\},\label{eq_cone_feasibleset}
\end{align}
with the functions $A_c:\mathbb{R}^{n_x}\rightarrow\mathbb{R}^{n_c \times n_u}$, $b_c:\mathbb{R}^{n_x}\rightarrow\mathbb{R}^{n_c}$ and $c_c:\mathbb{R}^{n_x}\rightarrow\mathbb{R}^{n_c}_+$ characterizing the (safety) constraints with $n_c$ being the number of constraints. Note that the QCQP problem \eqref{eq_nonLipschitz_Optimization} is reduced to a standard QP problem when $c_c(\dot)\equiv 0$.

To simplify the discussion, we consider the case in which each row of $A_c(x)$ is a unit vector, and each element of $c_c(x)$ takes values from the interval $[0,\bar{c}]$ with constant $0<\bar{c}<1$.

If the feasible set $\mathcal{U}_c(x)$ is nonempty and convex for each $x\in\mathcal{X}$, then the QCQP problem \eqref{eq_nonLipschitz_Optimization} always admits one unique solution, and accordingly, the resulting map from $x$ to $u$ can be represented by a function:
\begin{align}
u=\rho_c(x).\label{eq_def_varrho_c}
\end{align}

Here is an example of safety control. This example demonstrates that the function $\rho_c$ can be non-Lipschitz. Furthermore, even when $\rho_c$ is Lipschitz, its Lipschitz constant may vary depending on the configuration of the obstacles.

\begin{example}\label{example_non_Lipschitz}
Consider a mobile agent modeled by $\dot{x}=u$, with $x\in\mathbb{R}^2$ as the state and $u\in\mathbb{R}^2$ as the control input. With a nominal controller $\rho_0(x)\equiv[1;0]$, it is expected to safely steer the mobile agent through the gap between the two disc-type obstacles with the same radius $0<D\leq 1$ but centered at $o_1=[0;1]$ and $o_2=[0;-1]$, respectively. The scenario is illustrated in Figure \ref{figure_example_norm_of_solutions_original}.

This naturally results in a QP problem in the form of \eqref{eq_nonLipschitz_Optimization} with
\begin{subequations}\label{eq_example_nonlipshitz_Acbccc}
\begin{align}
A_c(x)&= -\left[\frac{(x-o_1)^T}{|x-o_1|};\frac{(x-o_2)^T}{|x-o_2|}\right],\\
b_c(x)&=\left[\frac{h_1(x)}{2|x-o_1|};\frac{h_2(x)}{2|x-o_2|}\right],~~c_c(x) \equiv 0
\end{align}    
\end{subequations}
for $x\in\mathcal{X}=\{x:|x-o_i|>0,i=1,2\}$, where $h_i(x)=|x-o_i|^2-D^2$ for $i=1,2$. 

The invariance of the set $\{x:h_i(x)\ge 0,i=1,2\}$ and thus the safety of the mobile agent is guaranteed by Nagumo's Theorem \cite{Blanchini-Miani-2015-Book} as $h_i$ satisfies $\partial h_i(x)/\partial x \neq 0$ and $\inf\{\nabla h_i(x)u:u\in\mathcal{U}_c(x)\}\geq 0$ for all $x\in\{x:h_i(x)=0,i=1,2\}$. Moreover, we can verify that the feasible set $\mathcal{U}_c(x)$ is nonempty and convex for $x\in\mathcal{X}$, and corresponding to the QP problem \eqref{eq_nonLipschitz_Optimization}, we can represent the resulting map from $x$ to $u$ with a function $\rho_c$ as in \eqref{eq_def_varrho_c}.

However, the function $\rho_c$ may be non-Lipschitz or admit a Lipschitz constant depending on the configuration of the obstacles. Indeed, for the special case of restricting $[x]_2=0$, by solving the above QP problem, we have
\begin{align}
\rho_c(x) =\begin{cases}
\Big[\frac{D^2-[x]_1^2-1}{2[x]_1}; 0\Big],&\text{for}~[x]_1\in(-D-1, D-1);\\
[1;0],&\text{for}~[x]_1\notin(-D-1, D-1).
\end{cases}
\end{align}
For $[x]_1\in(-D-1, D-1)$ and $[x]_2\equiv0$,
\begin{align}
\frac{\partial\rho_c(x)}{\partial [x]_1} = \frac{1-[x]_1^2-D^2}{2[x]_1^2}.\label{eq.example01.derivative}
\end{align}
The Lipschitz constant of $\rho_c$ when $[x]_2\equiv0$ is not less than $D/(1-D)$ if $0<D<1$, and $\rho_c$ is not Lipschitz at $x=0$ if $D=1$. 

\begin{figure}[!ht]
\centering
\includegraphics[width=0.75\linewidth]{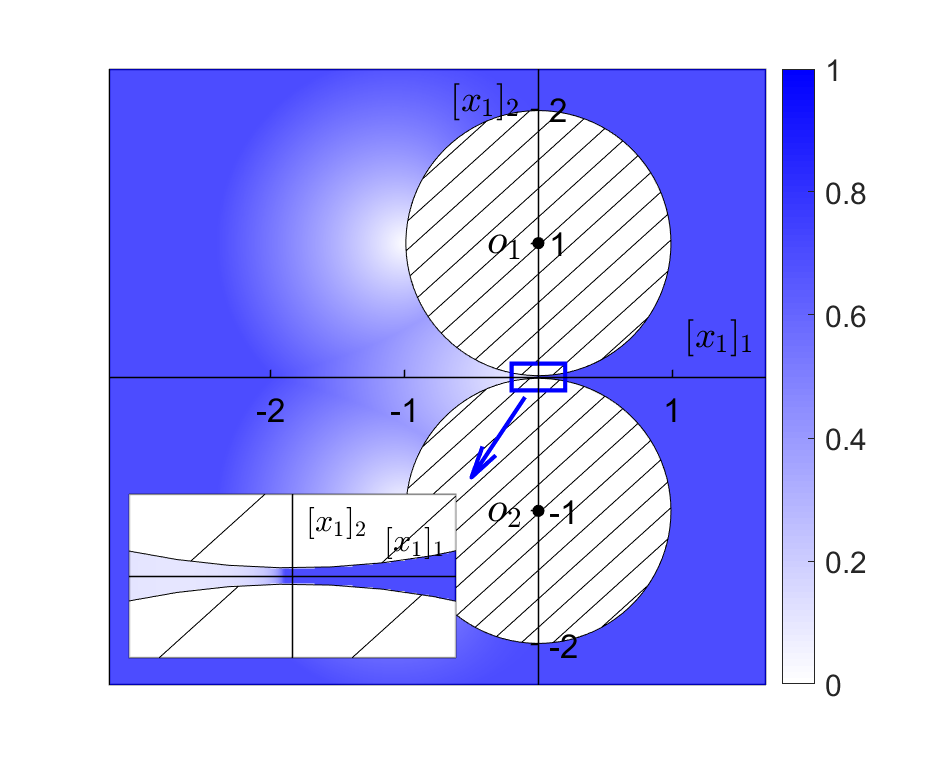}
\caption{$|\rho_c(x)|$ with respect to different $x$: $\rho_c$ may be non-Lipschitz or admit a Lipschitz constant depending on the configuration of the obstacles.}
\label{figure_example_norm_of_solutions_original}
\end{figure} 
\end{example}

\subsection{Feasible-Set Reshaping}
\label{subsection_feasible_set_reshaping_technique}

According to \cite[Theorem 3.1]{Hager-SIAMControl-1979} (also see Theorem \ref{theorem_hager} in the Appendix), the Lipschitz continuity of the solution to a QP problem with respect to the parameters (also called data in the literature of numerical optimization \cite{Hager-SIAMControl-1979}) requires a full-row rank coefficient matrix of the non-redundant active constraints. Based on this understanding, we reshape the original feasible set with an appropriately chosen positive basis \cite{Davis-AJM-1954}. The employment of the positive basis is motivated by the facts given by the following Lemmas \ref{lemma_reshape} and \ref{lemma_positive_basis}.

\begin{lemma} \label{lemma_reshape}
Consider a set-valued map $\mathcal{U}_c:\mathcal{X}\leadsto\mathbb{R}^{n_u}$ with a selection $\rho_s: \mathcal{X} \rightarrow \mathbb{R}^{n_u}$, i.e., $\rho_s(x)\in\mathcal{U}_c(x)$ for all $x\in\mathcal{X}$. For any $A_L = [l_1, \ldots, l_{n_l}]^T\in\mathbb{R}^{n_l\times n_u}$ with $l_{i}\in\mathbb{R}^{n_u}$ for $i=1,\ldots, n_l$ forming a positive basis, there exists a function $b_L: \mathcal{X} \rightarrow \mathbb{R}^{n_l}$ such that
\begin{align}
\mathcal{U}_L(x) := \{u\in \mathbb{R}^{n_u}: A_Lu\le b_L(x)\}  \label{eq_linear_feasible_set}
\end{align}
is non-empty and
\begin{align}
\rho_s(x) &\in \mathcal{U}_L(x) \subseteq \mathcal{U}_c(x), \label{eq_reshape_set_properties}
\end{align}
holds for all $x \in \mathcal{X}$. 
If, moreover, the function $\rho_s$ is (locally) Lipschitz, then the function $b_L$ can be chosen to be (locally) Lipschitz.
\end{lemma}

The proof of Lemma \ref{lemma_reshape} is placed in Appendix \ref{appendix_lemma_reshape_proof}.

\begin{lemma}\label{lemma_positive_basis}
For specific constant $c_A$ satisfying $0<c_A<1$, there exists a matrix $A_L = [l_1, \ldots, l_{n_l}]^T$ with $l_{i}\in\mathbb{R}^{n_u}$ for $i=1,\ldots, n_l$ and $n_l>n_u$ such that 
\begin{enumerate}
\item Each $l_i$ is a unit vector, i.e., $|l_i| = 1$;
\item Any $n_u$ vectors selected from $l_1,\ldots,l_{n_l}$ are linearly independent;
\item Any unit vector $l_0\in\mathbb{R}^{n_u}$ is a positive combination of the elements of $\{l_i : l_i^Tl_0 \ge c_A, i = 1,\ldots, n_l \}$, whose cardinality is not less than $n_u$.
\end{enumerate}
\end{lemma}

Lemma \ref{lemma_positive_basis} is basically a reformulation of the properties (37)--(38) in \cite{Wu-Liu-Egerstedt-Jiang-2023-TAC}.
See \cite{Wu-Liu-Egerstedt-Jiang-2023-TAC} for the proof.

Suppose that $\mathcal{U}_c:\mathcal{X}\rightarrow\mathbb{R}^{n_u}$ admits a selection $\rho_s:\mathcal{X}\rightarrow\mathbb{R}^{n_u}$. Define a set-valued map $\mathcal{U}_L:\mathcal{X}\leadsto\mathbb{R}^{n_u}$, referred to as the reshaped feasible set, in the form of \eqref{eq_linear_feasible_set}, where $A_L\in\mathbb{R}^{n_l\times n_u}$ is chosen to admit the properties given by Lemma \ref{lemma_positive_basis} with constant $c_A$, and $b_L:\mathcal{X}\rightarrow\mathbb{R}^{n_l}$ is a function satisfying \eqref{eq_reshape_set_properties}. Replacing the set-valued map $\mathcal{U}_c$ in \eqref{eq_nonLipschitz_Optimization} with $\mathcal{U}_L$ leads to
\begin{subequations}
\label{eq_safety_control_law_varrhoL}
\begin{align}
&u = \argmin_u |u-\rho_0(x)|^2 \label{eq_reshaped_control_law_cost_function}  \\
&\text{s.t.}~u\in\mathcal{U}_{L}(x).
\end{align}
\end{subequations}

Theorem \ref{theorem_reshaping} gives the properties of the QP problem \eqref{eq_safety_control_law_varrhoL} with the reshaped feasible set.

\begin{theorem}\label{theorem_reshaping}
Consider the QP problem \eqref{eq_safety_control_law_varrhoL} with $\mathcal{U}_L$ defined by $A_L$ and $b_L$ as described before \eqref{eq_safety_control_law_varrhoL}. 
Given the nominal controller $\rho_0$ and the QCQP problem \eqref{eq_nonLipschitz_Optimization} with $\mathcal{U}_c$ defined by \eqref{eq_cone_feasibleset}.
Suppose that Assumption \ref{assumption_nominal_controller} is satisfied, and $\mathcal{U}_c:\mathcal{X}\rightarrow\mathbb{R}^{n_u}$ admits a selection $\rho_s:\mathcal{X}\rightarrow\mathbb{R}^{n_u}$. Then, 
\begin{enumerate}
\item the QP problem \eqref{eq_safety_control_law_varrhoL} admits a unique solution for each $x\in\mathcal{X}$;
\item the function $\rho_L:\mathcal{X}\rightarrow\mathbb{R}^{n_u}$ representing the map from $x$ to $u$ generated by the QP problem \eqref{eq_safety_control_law_varrhoL} satisfies
\begin{align}
|\rho_L(x)| &\le |\rho_s(x)| + |\rho_0(x)| \label{eq_solution_upperbound}
\end{align}
for all $x\in\mathcal{X}$;
\item if $\rho_s$ and $b_L$ are Lipschitz on $\mathcal{X}$, then there exists a positive constant $L_{\rho_L}$ such that
\begin{align}
|\rho_L(x_1)-\rho_L(x_2)| \le L_{\rho_L}|x_1-x_2|.\label{eq_Lipschitz_property_of_qcqp}
\end{align}
for all $x_1,x_2\in\mathcal{X}$.
\end{enumerate}
\end{theorem}

The proof of Theorem \ref{theorem_reshaping} is given in Section \ref{subsection_proof_of_reshaping_theorem}.

The following proposition gives a practical method to realize the feasible-set reshaping technique.

\begin{proposition} \label{proposition_practical_reshape}
For $x\in\mathcal{X}$, given the original feasible set $\mathcal{U}_c(x)$ defined by \eqref{eq_cone_feasibleset} with $A_c$, $b_c$ and $c_c$, consider the reshaped feasible set $\mathcal{U}_L(x)$ in \eqref{eq_linear_feasible_set} defined by $A_L\in\mathbb{R}^{n_l\times n_u}$ and $b_L:\mathcal{X}\rightarrow\mathbb{R}^{n_l}$.
Suppose that $\mathcal{U}_c:\mathcal{X}\rightarrow\mathbb{R}^{n_u}$ admits a selection $\rho_s:\mathcal{X}\rightarrow\mathbb{R}^{n_u}$, and $A_L$ admits the properties given by Lemma \ref{lemma_positive_basis} with constant 
\begin{align}
c_A > \cos(\pi/2 - \acos (\sqrt{1-\bar{c}^2})). \label{eq_c_A_codition}
\end{align}
Then, property \eqref{eq_reshape_set_properties} holds for all $x\in\mathcal{X}$ by defining
\begin{align}
b_L(x)      &= \phi(\rho_s(x), A_c(x), b_c(x), c_c(x)) \label{eq_def_b_L}
\end{align}
where 
\begin{subequations} \label{eq_def_practical_reshape}
\begin{align}
\phi(s, A,b,c) &= A_Ls + \min\{\phi_1(s,A,b,c)+\phi_2(A)\} \\
\phi_1(s,A,b,c) &= \max(A_LA^T, \bar{c}_A)\odot\frac{(b-As-c|s|)^T}{1+c^T} \\
\phi_2(A) &= \max(k_{\phi}(\bar{c}_A-A_LA^T), 0)
\end{align}    
\end{subequations}
with $\phi:\mathbb{R}^{n_u}\times\mathbb{R}^{n_c\times n_u} \times \mathbb{R}^{n_c} \times \mathbb{R}^{n_c}_+ \rightarrow\mathbb{R}_{+}^{n_l}$, $\phi_1:\mathbb{R}^{n_u}\times\mathbb{R}^{n_c\times n_u} \times \mathbb{R}^{n_c} \times \mathbb{R}^{n_c}_+ \rightarrow\mathbb{R}_{+}^{n_l \times n_c}$, $\phi_2:\mathbb{R}^{n_c\times n_u}\rightarrow\mathbb{R}_{+}^{n_l \times n_c}$, $\bar{c}_A=\cos(\acos\sqrt{1-\bar{c}^2} + \acos c_A)$ and $k_{\phi}$ being any nonnegative constant. 
\end{proposition} 

See Appendix \ref{appendix_proof_practical_reshape} for the proof.

\begin{remark}
Following Proposition \ref{proposition_practical_reshape}, for any original feasible set $\mathcal{U}_c(x)$ with $x\in\mathcal{X}$, the process of reshaping the feasible set consists of the following steps:
\begin{enumerate}
\item Constructing a Lipschitz selection $\rho_s$ of the set-valued map $\mathcal{U}_c$;
\item Finding a specific positive basis $A_L$ in the $\mathbb{R}^{n_u}$ space;
\item Expanding with an appropriate range around the selection $\rho_s$, in the direction of each basis vector of the positive basis.
\end{enumerate}
See Lemma \ref{lemma_Lipschitz_selection} and Example \ref{example_reshape}.
\end{remark}

The following example shows how the feasible-set reshaping technique is used to address the non-Lipschitz issue in safety control.

\begin{example} \label{example_reshape}
Continue Example \ref{example_non_Lipschitz}. We apply the feasible-set reshaping technique given by Proposition \ref{proposition_practical_reshape} to the original feasible set $\mathcal{U}_c(x)$ defined by \eqref{eq_cone_feasibleset} with $A_c$, $b_c$ and $c_c$ defined by \eqref{eq_example_nonlipshitz_Acbccc}.
First, we can verify that $A_c$, $b_c$ and $c_c$ are Lipschitz continuous with respect to $x\in\mathcal{X}$, $\rho_s(x)=0$ for all $x\in\{x:|x-o_i|\ge D, i=1,2\}$, and condition \eqref{eq_Lipschitz_condition} is satisfied for $x\in\mathcal{X}$. 

Choosing $b_L$ according to \eqref{eq_def_b_L}, $c_A=\cos(2\pi/n_l)$ with $n_l=5$ and $k_{\phi}=0$, we can obtain $\mathcal{U}_L$ in the form of \eqref{eq_linear_feasible_set} with
\begin{align}
[A_L]_{j,:}&=[\cos(2\pi j/n_l), \sin (2\pi j/n_l)],~j=1,\ldots,n_l, \\
b_L(x) &= \min\{\max(A_LA_c^T(x), c_A)\odot b_c^T(x)\}.
\end{align}

Then, for $[x]_2=0$, when active constraints of the QP problem \eqref{eq_safety_control_law_varrhoL} present, we have 
\begin{align}
\rho_L(x) &= \Big[\max\Big(\frac{-[x]_1}{\sqrt{[x]_1^2+1}},c_A\Big)\frac{1+[x]_1^2-D^2}{2\sqrt{1+[x]_1^2}}; 0\Big].
\end{align}
This verifies the validity of inequality \eqref{eq_Lipschitz_property_of_qcqp}.

Figure \ref{figure_example_norm_of_solutions_reshaped} shows how $x$ influences $|\rho_L(x)|$ in the case of $D=0.99$ and $k_{\phi}=1.0$, which is in accordance with our theoretical analysis of Lipschitz continuity.
\begin{figure}[h]
\centering
\includegraphics[width=0.75\linewidth]{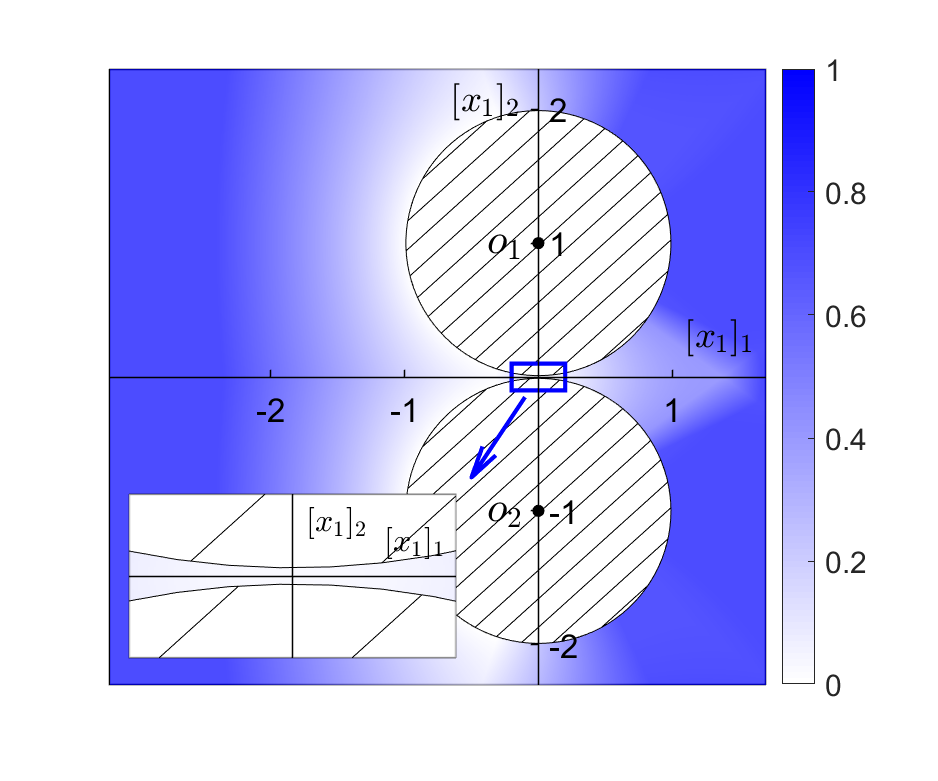}
\caption{$|\rho_L(x)|$ with respect to different $x$. In the scenario of Example \ref{example_non_Lipschitz}, the Lipschitz continuity of the solution to the QP problem \eqref{eq_safety_control_law_varrhoL} can be guaranteed by reshaping the feasible set.}
\label{figure_example_norm_of_solutions_reshaped}
\end{figure} 
\end{example}

\subsection{Proof of Theorem \ref{theorem_reshaping}}
\label{subsection_proof_of_reshaping_theorem}

\subsubsection{Property 1)}
With Lemma \ref{lemma_reshape}, each $x\in\mathcal{X}$, $\rho_s(x)\in\mathcal{U}_c(x)$ together with property $\rho_s(x) \in\mathcal{U}_L(x)$ guarantees that $\mathcal{U}_L(x)\neq \emptyset$.
It can be observed that $\mathcal{U}_L(x)$ is a convex set for each $x\in\mathcal{X}$.
Then, using projection theorem \cite[Proposition B.11]{Bertsekas-book-1997}, we conclude that the QP problem \eqref{eq_safety_control_law_varrhoL} admits a unique solution for each $x\in\mathcal{X}$. Property 1) is proved.

\subsubsection{Property 2)}
For each $x\in\mathcal{X}$, the projection theorem \cite[Proposition B.11]{Bertsekas-book-1997}, together with $\rho_s(x)\in\mathcal{U}_L(x)$, leads to the following inequality:
\begin{align}
(\rho_s(x)-\rho_L(x))^T(\rho_0(x)-\rho_L(x)) \le 0,
\end{align}
which equals 
\begin{align}
\left|\rho_L(x)-\frac{\rho_0(x)+\rho_s(x)}{2}\right|^2-\left|\frac{\rho_0(x)-\rho_s(x)}{2}\right|^2 \le 0. \label{eq_projection_steps_01}
\end{align}
Applying the triangle inequality \cite[Theorem 3.3]{Apostol-book-1973} to the inequality \eqref{eq_projection_steps_01} guarantees \eqref{eq_solution_upperbound} for all $x\in\mathcal{X}$. This ends the proof of Property 2).

\subsubsection{Property 3)}
We first prove the Lipschitz continuity of the QP-based safety controller \eqref{eq_safety_control_law_varrhoL} using Theorem \ref{theorem_hager} in Appendix \ref{appendix_hager}.
For $x\in\mathcal{X}$, define a new QP problem 
\begin{subequations}\label{eq_new_QP}
\begin{align}
&u' = \argmin_u |u-\rho_0'(x)|^2 \\
&\text{s.t. } u\in\mathcal{U}'_L(x).
\end{align}    
\end{subequations}
where $\rho_0'(x)=\rho_0(x)-\rho_s(x)$ and $\mathcal{U}'_L(x)=\{u:A_L u \le b'_L(x)\}$ with $b'_L(x) = b_L(x)-A_L\rho_s(x)$.
It can be verified that $\mathcal{U}'_L(x)\neq \emptyset$ by using $\rho_s(x)\in\mathcal{U}_L(x)$. 
Then, $\rho_L$ is equivalent to
\begin{align}
\rho_L(x) = \rho_s(x)+ u', \label{eq.equivalent_form}
\end{align}
where $u'$ is the solution to QP problem \eqref{eq_new_QP}.

Define $\breve{A}_L$ and $\breve{b}'_L(x)$ as the non-redundant active constraints of the QP problem \eqref{eq_new_QP}. 
For any $x\in\mathcal{X}$, $\rho'_r(x)$, $A_L$ and $b'_L(x)$ are elements of $\mathbb{R}^{n_u}$, $\{A_L\}$ and $\mathbb{R}^{n_l}_+$, respectively, and all the conditions required by Theorem \ref{theorem_hager} given in Appendix \ref{appendix_hager} are checked as follows:
\begin{itemize}
\item $\mathbb{R}^{n_u}$, $\{A_L\}$ and $\mathbb{R}^{n_l}_+$ are convex;
\item for each $x\in\mathcal{X}$, $0\in\mathcal{U}'_L(x)$ and the projection theorem \cite[Proposition B.11]{Bertsekas-book-1997} guarantee the existence and uniqueness of the solution to QP problem \eqref{eq.equivalent_form}, respectively;
\item the first property given by Lemma \ref{lemma_positive_basis} guarantees the boundedness of the norm of $\breve{A}_L$;
\item the second property given by Lemma \ref{lemma_positive_basis} guarantees the existence of a constant $c_{\lambda}>0$ such that $\lambda_{\min}^{1/2}(\breve{A}_L\breve{A}_L^T) \ge c_{\lambda}$, and thus $|\breve{A}_L^T\xi|\ge c_{\lambda}|\xi|$ for all $\xi$.
\end{itemize}
Then, the solution to the QP problem \eqref{eq_new_QP} is Lipschitz continuous with respect to $\rho_0'(x)$ and $b_L'(x)$, and thus with respect to $x\in\mathcal{X}$. 

Using inequality \eqref{eq_varrhoL_Lipschitz}, there exists a positive constant $L$ such that for any $x_1$, $x_2\in\mathcal{X}$,
\begin{align}
&|\rho_L(x_1)-\rho_L(x_2)| \le |\rho_s(x_1)-\rho_s(x_2)| \notag \\
&~~~+ L(|\rho_0'(x_1)-\rho_0'(x_2)|+|b'_L(x_1)-b'_L(x_2)|) \notag \\
&\le (1+L(1+\lambda_{\max}(A_L^TA_L)))|\rho_s(x_1)-\rho_s(x_2)| \notag \\
&~~~+ L(|\rho_0(x_1)-\rho_0(x_2)|+|b_L(x_1)-b_L(x_2)|), \label{eq_qp_is_Lispchitz}
\end{align}
where we have used the definitions of $\rho'_0(x)$ and $b'_L(x)$ after \eqref{eq_new_QP} for the last inequality.

By Assumption \ref{assumption_nominal_controller}, $\rho_0$ is Lipschitz continuous on $\mathcal{X}$. Inequality \eqref{eq_qp_is_Lispchitz}, together with the Lipschitz continuity of $\rho_0$, $\rho_s$ and $b_L$, guarantees \eqref{eq_Lipschitz_property_of_qcqp}. This ends the proof of Property 3).

\section{Constructive Safety Control of Nonlinear Plants in the Cascade Form}
\label{section_main_result}

With the design ingredients developed in Sections \ref{section_relative_degree_one} and \ref{section_feasible_set_reshaping}, this section proposes the main result of this paper by means of a refined nonlinear small-gain synthesis. 

The constructive controller design process is implemented by recursively applying the safety control and feasible-set reshaping techniques introduced in Sections \ref{section_relative_degree_one} and \ref{section_feasible_set_reshaping}.
The recursive design results in an interconnected system composed of the controlled subsystems of the plant, for which a nonlinear small-gain synthesis is employed to fine-tune the controller parameters to ensure the safety of the closed-loop system.

\subsection{Initial Step: Safety Control of the $x_1$-Subsystem}

Rewrite the $x_1$-subsystem given by \eqref{eq_plant_dynamics} with $i=1$ as 
\begin{align}
\dot{x}_1 = f_1(z_1)+(g_1+\delta_1(z_1))(x^*_2+\tilde{x}_2), \label{eq_x1subsystem_rewrite}
\end{align}
where $x^*_2\in\mathbb{R}^{n_{x_2}}$ is considered as the virtual control input and $\tilde{x}_2$ is the new error state defined by 
\begin{align}
\tilde{x}_2 = x^*_2-x_2.
\end{align}

For the $x_1$-subsystem \eqref{eq_x1subsystem_rewrite}, motivated by the QCQP algorithm \eqref{eq_safety_control_law} and the feasible-set reshaping technique \eqref{eq_safety_control_law_varrhoL}, we design a virtual control law 
\begin{subequations} \label{eq_safety_controller}
\begin{align}
& x^*_2 = \argmin_u |u-\rho_0(x_1)|^2 \\
& \text{s.t.~~} u\in\mathcal{U}_1(x_1), 
\end{align}    
\end{subequations}
where $\rho_0$ is the nominal controller satisfying Assumption \ref{assumption_nominal_controller}, and $\mathcal{U}_1(x_1)$ is the reshaped feasible set defined by
\begin{align}
\mathcal{U}_1(x_1) = \{u\in\mathbb{R}^{n_{x_2}}:A_Lu\le b_L(x_1)\}
\end{align}
with $A_L=[l_1,\ldots, l_{n_l}]^T$ chosen to admit the properties given by Lemma \ref{lemma_positive_basis} with constant $c_A$ satisfying \eqref{eq_c_A_codition} and $b_L:\mathcal{X}\rightarrow \mathbb{R}^{n_l}$ defined by \eqref{eq_def_b_L} with 
\begin{subequations}
\begin{align}
[A_c(x_1)]_{j,:} &= \frac{\partial V_j(x_1)}{\partial x_1} g_1\Big|\frac{\partial V_j(x_1)}{\partial x_1} g_1\Big|^{-1}, \label{eq_def_highorder_Ac}\\
[b_c(x_1)]_j &= -\alpha_j(V_j(x_1)-v_j),\label{eq_def_highorder_bc}\\
[c_c(x_1)]_j &= \bar{\delta} \ubar{g}^{-1}, \label{eq_def_highorder_cc}
\end{align}    
\end{subequations}
for $j=1,\ldots,n_c$.

As a direct application of Theorems \ref{theorem_safety_relative_degree_one} and \ref{theorem_reshaping}, the following proposition shows the ISpSf property of the controlled $x_1$-subsystem.

\begin{proposition}\label{proposition_outer_loop}
Given the nominal controller $\rho_0$, a function $\gamma_{1,2}\in\mathcal{K}$, the certificate functions $V_j$ and $c_j>v_j$ for $j=1,\ldots,n_c$, consider the $x_1$-subsystem defined by \eqref{eq_x1subsystem_rewrite} with $i=1$ and the safety control law \eqref{eq_safety_controller}. Suppose that Assumptions \ref{assumption_high_relative_degree}, \ref{assumption_certificate_functions} and \ref{assumption_nominal_controller} hold, and conditions \eqref{eq_ubarg_bardelta}--\eqref{eq_chosen_of_bc} in Theorem \ref{theorem_safety_relative_degree_one} are satisfied.
Then, 
\begin{enumerate}
\item the QP problem \eqref{eq_safety_controller} admits a unique solution for all $x\in\mathcal{X}$, and the resulting map from $x_1$ to $x^*_2$ can be represented by a function $x^*_2=\rho_1(x_1)$;
\item for $j=1,\ldots,n_c$ and any piecewise continuous $\tilde{x}_2$, along the trajectories of the $x_1$-subsystem \eqref{eq_highorder_model} and \eqref{eq_safety_controller}, each safety certificate function $V_j:\mathcal{X}\rightarrow\mathbb{R}_+$ satisfies
\begin{align}
\!\!\!\!V_j(x_1(t)) \le \max\{\beta_1(V_j(x_1(0)), t),\gamma_{1,2}(\|\tilde{x}_2\|_t), c_j\}\!\! \label{eq_ISpS_high_relative_degree}
\end{align}
whenever $x_1$ and $\tilde{x}_2$ are defined, with $\beta_1(s,t)=s e^{-\eta\ubar{g}\theta t}$;
\item there exists a function $\gamma_{x^*_2,V}\in\mathcal{K}$ such that
\begin{align}
|x^*_2|\le&\gamma_{x^*_2,V}\big(\breve{V}(x_1)\big) + \bar{\rho}_0, \label{eq_bound_outer_controllaw}
\end{align}
for any $x_1\in\mathcal{X}$, and moreover, if $A_c$ and $b_c$ are Lipschitz on $\mathcal{X}$, then, for any $x_1\in\mathcal{X}$ and any piecewise continuous $\tilde{x}_2$, there exists a positive constant $k_1$ such that 
\begin{align}
|D^+x^*_2(t)| &\le k_1 (\bar{f}_{x_1}(|x_1(t)|) \notag \\
&~~~+(\bar{g}+\bar{\delta})(|x^*_2(t)|+|\tilde{x}_2(t)|)), \label{eq_Lipschitz_outer_controllaw}
\end{align}
whenever $x_1$ and $\tilde{x}_2$ are defined.
\end{enumerate}
\end{proposition}

\begin{IEEEproof}
{\it Property 1):}
Consider the set-valued map $\mathcal{F}_c$ defined by \eqref{eq_practical_control_input_set} with $n_u=n_{x_2}$, and the functions $A_c$ and $b_c$ defined by \eqref{eq_def_highorder_Ac} and \eqref{eq_def_highorder_bc}, respectively. It can be verified that $\mathcal{F}_c(x_1)\neq \emptyset$ for all $x_1\in\mathcal{X}$ based on property 1) of Theorem \ref{theorem_safety_relative_degree_one}. 
Also, using Proposition \ref{proposition_practical_reshape}, we have 
\begin{align}
\mathcal{U}_1(x_1)\neq \emptyset \text{~~~and~~~} \mathcal{U}_1(x_1) \subseteq \mathcal{F}_c(x_1)
\end{align}
for all $x_1\in\mathcal{X}$. Then, property 1) of Proposition \ref{proposition_practical_reshape} can be proved with property 1) of Theorem \ref{theorem_reshaping}.

{\it Property 2):}
For each $V_1, \ldots, V_{n_c}$, the validity of this property can be readily confirmed by applying property 2) of Theorem \ref{theorem_safety_relative_degree_one}.
In particular, $x_1$, $0$, $\rho_1$, $\tilde{x}_2$, $\gamma_{1,2}$, and $\beta_1$ in Proposition \ref{proposition_outer_loop} correspond to $x$, $z$, $\rho$, $w$, $\gamma^w$, and $\beta$ in Theorem \ref{theorem_safety_relative_degree_one}, respectively.

{\it Property 3):} We first consider inequality \eqref{eq_bound_outer_controllaw}.

Following the analysis below \eqref{eq_certificate_distance} in the proof of Theorem \ref{theorem_safety_relative_degree_one}, it can be verified that condition \eqref{eq_Lipschitz_condition} required by Lemma \ref{lemma_Lipschitz_selection} is satisfied. The property \eqref{eq_selection_of_Uc} of Lemma \ref{lemma_Lipschitz_selection} directly guarantees that $\mathcal{U}_c(x_1)\neq\emptyset$ for all $x\in\mathcal{X}$, which is required by Theorem \ref{theorem_reshaping}.

 Then, using the property \eqref{eq_selection_of_Uc} of Lemma \ref{lemma_Lipschitz_selection} and property 2) of Theorem \ref{theorem_reshaping}, we have 
\begin{align}
|x^*_2| &\le |\rho_s(x_1)| + |\rho_0(x_1)| \notag \\
&\le \left|\min\left\{0, \min\limits_{j=1,\ldots,n_c} [b_c(x_1)]_j\right\}\frac{\ubar{g}}{\ubar{g}-\bar{\delta}}\right| + |\rho_0(x_1)| \notag \\
&\le \frac{\max\{\alpha_{\bar{j}}(V_{\bar{j}}(x_1)-v_{\bar{j}}),0\}}{1-\bar{\delta}/\ubar{g}} + \bar{\rho}_0, 
\end{align}
where we have used property 2) of Theorem \ref{theorem_reshaping} for the first inequality, used the definition of $\rho_s$ in \eqref{eq_selection} for the second inequality, and used the definition of $b_c$ in \eqref{eq_def_highorder_bc} and Assumption \ref{assumption_nominal_controller} for the third inequality, with $\bar{j}$ representing the index of any one of the largest elements of $V_j(x_1)$ for $j=1,\ldots, n_c$.
This guarantees the existence of $\gamma_{x^*_2,V}\in\mathcal{K}$ for the satisfaction of \eqref{eq_bound_outer_controllaw}.

Inequality \eqref{eq_Lipschitz_outer_controllaw} is proved by using property 3) of Theorem \ref{theorem_reshaping}, which requires that $\rho_s$ and $b_L$ are Lipschitz on $\mathcal{X}$.
The Lipschitz continuity of $b_L$ is verified by the Lipschitz continuity of $A_c$ and $b_c$, as well as the definition of $b_L$ in \eqref{eq_def_b_L}. The Lipschitz continuity of $\rho_s$ is guaranteed by Lemma \ref{lemma_Lipschitz_selection}.
Then, using property 3) of Theorem \ref{theorem_reshaping}, we can prove the existence of a positive constant $k_1$ such that \eqref{eq_Lipschitz_outer_controllaw} holds whenever $x_1$ and $\tilde{x}_2$ are defined. This ends the proof of Proposition \ref{proposition_outer_loop}.
\end{IEEEproof}

\subsection{Recursive Steps: Constrained Tracking Control of the $x_i$-Subsystems ($i=2,\ldots,m$)}

Following the convention of constructive nonlinear control, rewrite the $x_i$-subsystem \eqref{eq_plant_dynamics} as 
\begin{align}
\dot{x}_i = f_i(z_i)+(g_i+\delta_i(z_i))(x^*_{i+1}+\tilde{x}_{i+1}),i=2,\ldots,m \!\label{eq_highorder_model} 
\end{align}
where $x^*_{i+1}\in\mathbb{R}^{n_{x_{i+1}}}$ is considered as the virtual control input and $\tilde{x}_{i+1}$ is considered as the new error state defined by
\begin{align}
\tilde{x}_{i+1} = x^*_{i+1}-x_{i+1}, \label{eq_tracking_error}
\end{align}
for $i=1,\ldots,m$, with $x_{m+1}=u$. 

The upper right Dini derivative of $\tilde{x}_i$ is given by:
\begin{align}
D^+\tilde{x}_i &= f_i(z_i) + (g_i + \delta_i(z_i))(x^*_{i+1}+\tilde{x}_{i+1}) - D^+x^*_i, \label{eq_dini_tilde_x_i}
\end{align}
where $x^*_{i+1}\in\mathbb{R}^{n_{x_{i+1}}}$ is the virtual control input, and $\tilde{x}_{i+1}\in\mathbb{R}^{n_{x_{i+1}}}$, $D^+x^*_i\in\mathbb{R}^{n_{x_i}}$ and $z_i\in\mathbb{R}^{n_{z_i}}$ are considered as disturbances.

Now, we apply the design proposed in Section \ref{section_relative_degree_one} to the $\tilde{x}_i$-subsystem \eqref{eq_dini_tilde_x_i}. 
The dynamics of the $\tilde{x}_i$-subsystem is rewritten as: 
\begin{align}
D^+\tilde{x}_i&= f_i'(z_i', \tilde{x}_i) + (g_i + \delta'_i(z_i', \tilde{x}_i))(x^*_{i+1}+\tilde{x}_{i+1}),
\label{eq_dini_tilde_x_i_new_form}
\end{align}
with 
\begin{subequations}
\begin{align}
z_i' &= [z_{i-1}; x^*_i; D^+x^*_i],\\
f_i'(z_i', \tilde{x}_i) &= f_i([z_{i-1}; x^*_i+\tilde{x}_i]) - D^+x^*_i,\\
\delta_i'(z_i', \tilde{x}_i) &= \delta_i([z_{i-1}; x^*_i+\tilde{x}_i]).
\end{align}    
\end{subequations}
This, together with Assumption \ref{assumption_high_relative_degree}, ensures that the $\tilde{x}_i$-subsystem satisfies Assumption \ref{assumption_model_relative_degree_one}.
We consider a candidate Lyapunov function 
\begin{align}
V_{\tilde{x}_i}(\tilde{x}_i)=|\tilde{x}_i|,
\end{align}
which, if considered as a special safety certificate function, satisfies Assumption \ref{assumption_certificate_functions}. 

Following the design in Section \ref{section_relative_degree_one}, we propose the following virtual control law:
\begin{subequations} \label{eq_tracking_control_optimization}
\begin{align}
&x^*_{i+1} = \argmin_u |u|^2 \\
&\text{s.t.~} \frac{g^T_i\tilde{x}_i}{|g^T_i\tilde{x}_i|}u+\frac{\bar{\delta}}{\ubar{g}}|u| \le -\kappa_i(|\tilde{x}_i|),
\end{align}    
\end{subequations}
where $\kappa_i\in\mathcal{K}$ is to be determined later.

Under Assumption \ref{assumption_high_relative_degree} and $\ubar{g}>\bar{\delta}$, one may verify that the QCQP problem \eqref{eq_tracking_control_optimization} is feasible and admits a unique solution for each $\tilde{x}_i\neq 0$, and thus, directly represent the control law \eqref{eq_tracking_control_optimization} as:
\begin{align}
x^*_{i+1} &= \rho_i(\tilde{x}_i), \label{eq_reference_tracking_controller}
\end{align}
with $\rho_i:\mathbb{R}^{n_{x_i}}\rightarrow\mathbb{R}^{n_{x_{i+1}}}$. Indeed, $\rho_i$ is explicitly given by
\begin{align}
\rho_i(\tilde{x}_i) = 
\begin{cases}
-\frac{g^T_i\tilde{x}_i}{|g^T_i\tilde{x}_i|} \frac{\ubar{g}}{\ubar{g}-\bar{\delta}} \kappa_i(|\tilde{x}_i|), &\text{for~} \tilde{x}_i \neq 0,\\
0,&\text{otherwise}.
\end{cases} \label{eq_reference_tracking_control_law}
\end{align}

As an application of Theorem \ref{theorem_safety_relative_degree_one}, Proposition \ref{proposition_tracking_controller} shows the reference-tracking capability of each controlled $x_i$-subsystem ($i=2,\ldots,m$).

\begin{proposition} \label{proposition_tracking_controller}
Given any specific constant $\theta$ and functions $\gamma_{i,x_1},\gamma_{i,\rho_0},\gamma_{i,1},\ldots, \gamma_{i,i-1},\gamma_{i,i+1}\in\mathcal{K}\cup\{0\}$, for $i=2,\ldots, m$, consider the $\tilde{x}_i$-subsystem \eqref{eq_dini_tilde_x_i} and control law $\rho_i$ defined by \eqref{eq_reference_tracking_control_law}. 
Suppose that Assumption \ref{assumption_high_relative_degree} and $\ubar{g}>\bar{\delta}$ are satisfied, and there exist a constant $c_{x_1}\ge 0$ such that 
\begin{align}
\bar{f}_{x_1}(|x_1|) \le c_{x_1}, \label{eq_bound_f_x_1}
\end{align}
a continuously differentiable, odd, strictly increasing, radially unbounded function $\kappa_i$ and a constant $k_i$ such that 
\begin{align}
\kappa_i(s)&\ge \theta s + \frac{\ubar{g}+\bar{\delta}}{\ubar{g}}\gamma_{i,i+1}^{-1}(s) + \frac{1}{\ubar{g}} (\alpha_{i,x_1}\circ\gamma_{i,x_1}^{-1}(s)\notag \\ 
&~+\alpha_{i,\rho_0}\circ\gamma^{-1}_{i,\rho_0}(s) + \sum^{i-1}_{j=1}\alpha_{i,j}\circ\gamma_{i,j}^{-1}(s) + \alpha_{i,i}(s)) \label{eq_alpha_i_condition}\\
k_i &\ge \frac{\ubar{g}}{\ubar{g}-\bar{\delta}} \left(\frac{\bar{g}}{\ubar{g}}\frac{\kappa_i(s)}{s}+\frac{\mathrm{d} \kappa_i(s)}{\mathrm{d} s}\right) \label{eq_def_k_i}
\end{align}
for all $s\ge 0$, where 
\begin{subequations} \label{eq_tracking_gains}
\begin{align}
&\alpha_{i,x_1}(s) = \breve{k}_{1,i-1} s, \\
&\alpha_{i,\rho_0}(s) = \breve{k}_{2,i-1}\bar{f}_{x_2}(4s)+\bar{k}_{1,i-1}(\bar{g}+\bar{\delta})s, \\
&\alpha_{i,1}(s) = \alpha_{i,\rho_0}\circ \gamma_{x^*_2,V}(s), \\
&\alpha_{i,j}(s) = \breve{k}_{j,i-1}\bar{f}_{x_{j}}(2s) + \breve{k}_{j,i-1}\bar{f}_{x_{j+1}}(\frac{2\ubar{g} \kappa_j(s)}{\ubar{g}-\bar{\delta}}) + \notag \\
&(\bar{g}+\bar{\delta})(\bar{k}_{j,i-1}\frac{\ubar{g}\kappa_j(s)}{\ubar{g}-\bar{\delta}}+\bar{k}_{j-1,i-1}s), j=2,\ldots,i-1, \\
&\alpha_{i,i}(s) = \bar{f}_{x_i}(2s) + k_{i-1}(\bar{g}+\bar{\delta})s,
\end{align}   
\end{subequations}
with $\bar{k}_{p,i}=\Pi^{i}_{j=p}k_j$ and $\breve{k}_{p,i}=1+\sum^{i}_{j=p}\bar{k}_{j,i}$ (define $\bar{k}_{p,i}=0$ for any $p>i$).
Then, for any specific initial state $\tilde{x}_i(0)$, the tracking error $\tilde{x}_i(t)$ staisfies
\begin{align}
|\tilde{x}_i(t)|\le \max\{&\beta_i(|\tilde{x}_i(0)|, t), \gamma_{i,1}(\|\breve{V}(x_1)\|_t),\gamma_{i,2}(\Vert \tilde{x}_2 \Vert_t),\notag \\
&\ldots, \gamma_{i,i-1}(\|\tilde{x}_{i-1}\|_t), \gamma_{i,i+1}(\|\tilde{x}_{i+1}\|_t), \notag \\
&\gamma_{i,\rho_0}(\bar{\rho}_0), \gamma_{i,x_1}(c_{x_1}) \} \label{eq_x_tilde_final_traj}
\end{align}
whenever the variables are defined, with $\beta_i(s,t)=s e^{-\ubar{g}\theta t}$.
\end{proposition} 

\begin{IEEEproof}
Following the proof of Theorem \ref{theorem_safety_relative_degree_one}, taking the upper right Dini derivative of $|\tilde{x}_i|$ along the trajectories of the $\tilde{x}_i$-subsystem \eqref{eq_dini_tilde_x_i} and \eqref{eq_reference_tracking_control_law}, we have 
\begin{align}
D^+|\tilde{x}_i| \le \frac{|\tilde{x}_i^Tg_i|}{|\tilde{x}_i|}&\left(-\kappa_i(|\tilde{x}_i|) + \frac{\ubar{g}+\bar{\delta}}{\ubar{g}}|\tilde{x}_{i+1}|\right.\notag \\
&+\left.\frac{1}{\ubar{g}}(|f_i(z_i)|+|D^+x^*_i|)\right). \label{eq_dini_normxtilde_i}
\end{align}

To estimate the influence of $|f_i(z_i)|+|D^+x^*_i|$ in \eqref{eq_dini_normxtilde_i}, we first calculate the upper bound of $|D^+x^*_i|$.
For each $i=2,\ldots,m$, based on \eqref{eq_dini_tilde_x_i}, \eqref{eq_reference_tracking_controller} and \eqref{eq_reference_tracking_control_law}, it holds that
\begin{align}
|D^+x^*_{i+1}| &\le \bar{k}_{2,i}|D^+x^*_2| + \sum^{i}_{j=2}\bar{k}_{j,i}|f_j(z_j)|  \notag \\
&\hspace{-40pt} + (\bar{g}+\bar{\delta})\sum^{i}_{j=2}\bar{k}_{j,i}|\tilde{x}_{j+1}| + \frac{\ubar{g}(\bar{g}+\bar{\delta})}{\ubar{g}-\bar{\delta}} \sum^{i}_{j=2} \bar{k}_{j,i}\kappa_j(|\tilde{x}_j|)  \label{eq_def_dini_x_i_ref}
\end{align}
whenever the variables are defined. The validity of \eqref{eq_def_dini_x_i_ref} will be proved later.

Then, using condition \eqref{eq_assume_f_max_highorder}, definition \eqref{eq_tracking_error}, and properties \eqref{eq_bound_outer_controllaw}, \eqref{eq_Lipschitz_outer_controllaw} and \eqref{eq_def_dini_x_i_ref}, we have 
\begin{align}
&|f_i(z_i)|+ |D^+x^*_i| \le \alpha_{i,i}(|\tilde{x}_i|) + \alpha_{i,x_1}(c_{x_1})+ \alpha_{i,\rho_0}(\bar{\rho}_0)+ \notag \\
&~~~ \alpha_{i,1}(\breve{V}(x_1))+ \alpha_{i,2}(|\tilde{x}_2|)+ \cdots+ \alpha_{i,i-1}(|\tilde{x}_{i-1}|). \label{eq_norm_f_dinixref}
\end{align}
Substituting \eqref{eq_norm_f_dinixref} and \eqref{eq_alpha_i_condition} into \eqref{eq_dini_normxtilde_i}, we have
\begin{align}
&|\tilde{x}_i| \ge \max\{\gamma_{i,x_1}(c_{x_1}), \gamma_{i,\rho_0}(\bar{\rho}_0), \gamma_{i,1}(\breve{V}(x_1)),\gamma_{i,2}(|\tilde{x}_2|),\notag \\
&\ldots, \gamma_{i,i-1}(|\tilde{x}_{i-1}|), \gamma_{i,i+1}(|\tilde{x}_{i+1}|)\} \notag \\
\Rightarrow& D^+|\tilde{x}_i| \le -\ubar{g}\theta|\tilde{x}_i|.
\end{align}
Based on the discussion above, using the comparison principle \cite{Khalil-book-2002}, we can conclude that property \eqref{eq_x_tilde_final_traj} holds with $\beta_i(s,t)= s e^{-\theta\ubar{g}t}$ \cite{Sontag-2008}. 

Now we prove the validity of \eqref{eq_def_dini_x_i_ref}. Taking the upper right Dini derivative of $x^*_{i+1}$, we have 
\begin{align}
D^+x^*_{i+1} =& \frac{\ubar{g}}{\ubar{g}-\bar{\delta}} \left(I-\frac{g_i^T\tilde{x}_i\tilde{x}^T_ig_i}{|g^T_i\tilde{x}_i|^2}\right)\frac{g^T_i}{|g^T_i\tilde{x}_i|}\kappa_i(|\tilde{x}_i|)(D^+\tilde{x}_{i}) \notag \\
&+ \frac{\ubar{g}}{\ubar{g}-\bar{\delta}}\frac{\mathrm{d} \kappa_i}{\mathrm{d} |\tilde{x}_i|}\frac{g_i^T\tilde{x}_i}{|g_i^T\tilde{x}_i|}\frac{\tilde{x}_i^T}{|\tilde{x}_i|}(D^+\tilde{x}_{i}), 
\end{align}
and thus
\begin{align}
|D^+x^*_{i+1}| &\le \frac{\ubar{g}}{\ubar{g}-\bar{\delta}} \left(\left|\frac{g_i^T}{|g^T_i\tilde{x}_i|}\right|\kappa_i(|\tilde{x}_i|) + \frac{\mathrm{d} \kappa_i(|\tilde{x}_i|)}{\mathrm{d} |\tilde{x}_i|}\right)|D^+\tilde{x}_i|\notag\\
&\le \frac{\ubar{g}}{\ubar{g}-\bar{\delta}}\left(\frac{\bar{g}}{\ubar{g}} \frac{\alpha_i(|\tilde{x}_i|)}{|\tilde{x}_i|} +\frac{\mathrm{d} \alpha_i(|\tilde{x}_i|)}{\mathrm{d} |\tilde{x}_i|}\right)|D^+\tilde{x}_i| \notag \\
&\le k_i |D^+\tilde{x}_i| \label{eq_upper_bound_Dini_x_star_i_add_1}
\end{align}
where we have used the triangle inequality \cite[Theorem 3.3]{Apostol-book-1973} and the fact that the spectral radius of any idempotent matrix is $1$ for the first inequality, used the \eqref{eq_assume_g_min_highorder} and \eqref{eq_assume_g_max_highorder} for the second inequality, and used \eqref{eq_def_k_i} for the last inequality. 
Based on the upper right Dini derivative of $\tilde{x}_{i}$ defined in \eqref{eq_dini_tilde_x_i}, direct calculation yields
\begin{align}
|D^+\tilde{x}_i| &\le |D^+x^*_i| + \bar{f}_{x_1}(|x_1|) +\bar{f}_{z}(|\bar{z}_i|)  \notag \\
&~~~+ (\bar{g} + \bar{\delta})(|x^*_{i+1}|+|\tilde{x}_{i+1}|).
\end{align}
This, together with properties \eqref{eq_upper_bound_Dini_x_star_i_add_1}, \eqref{eq_reference_tracking_control_law} and \eqref{eq_reference_tracking_controller}, verifies the validity of \eqref{eq_def_dini_x_i_ref}. This ends the proof of Proposition \ref{proposition_tracking_controller}.
\end{IEEEproof}

\subsection{Small-Gain Synthesis and Main Result}
\label{subsection_main_result_small_gain}

The above-mentioned recursive design transforms the closed-loop system into an interconnected system, with the interaction between the subsystems described by gain functions, as shown in Figure \ref{theorem_small_gain}. 
\begin{figure}[!ht]
\centering
\begin{tikzpicture} [scale=1.0, transform shape]
\node at (-5.7,0) {$\breve{V}(x_1)$};
\node at (-3.7,0) {$\tilde{x}_2$};
\node at (-1.1,0) {$\tilde{x}_{i}$};

\node at (1.7,0) {$\tilde{x}_{m}$};
\node at (-2.4,0) {$\cdots$};
\node at (0.3,0) {$\cdots$};

\draw[->,>=stealth] (-4,0) -- (-5.1,0);
\draw[->,>=stealth] (-2.7,0) -- (-3.4,0);
\draw[->,>=stealth] (-1.4,0) -- (-2.1,0);

\draw[->,>=stealth] (-0.1,0) -- (-0.8,0);
\draw[->,>=stealth] (1.3,0) -- (0.6,0);

% x_i
\draw[->,>=stealth]  plot[smooth, tension=1.2] coordinates {(-3.4,0.2) (-2.4,0.5) (-1.3,0.2)};

% x1 gains
\draw[->,>=stealth]  plot[smooth, tension=1.2] coordinates {(-5.1,-0.2) (-3.2,-0.5) (-1.2,-0.2)};
\draw[->,>=stealth]  plot[smooth, tension=1.2] coordinates {(-5.1,0.2) (-4.5,0.4) (-3.9,0.2)};

\draw[->,>=stealth]  plot[smooth, tension=1.2] coordinates {(-5.2,-0.3) (-2,-0.9) (1.3,-0.2)};

% x_m gains
\draw[->,>=stealth]  plot[smooth, tension=1.5] coordinates { (-1,0.2) (0.1,0.5) (1.3,0.2)};
\draw[->,>=stealth]  plot[smooth, tension=1.5] coordinates { (-3.6,0.2) (-1.2,0.9) (1.4,0.3)};
\end{tikzpicture}
\caption{The closed-loop system as an interconnected system.}
\label{figure_ith_relationship}
\end{figure}
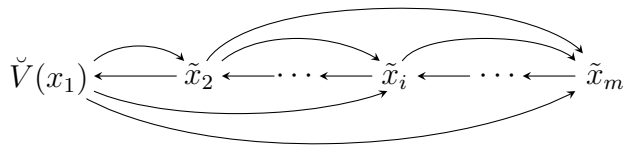

Through a nonlinear small-gain synthesis, we are able to fine-tune the safety controller such that the closed-loop system admits desired safety properties. 
% conditions \eqref{eq_alpha_j_definition}, \eqref{eq_feasibility_condition}, \eqref{eq_chosen_of_bc}, \eqref{eq_bound_f_x_1}, \eqref{eq_alpha_i_condition} and \eqref{eq_def_k_i}

\begin{theorem} \label{theorem_small_gain}
Consider the plant \eqref{eq_plant_dynamics}, the controller \eqref{eq_safety_controller} and \eqref{eq_reference_tracking_control_law}, and the certificate functions $V_1,\ldots, V_{n_c}$. Suppose that Assumptions \ref{assumption_certificate_functions}, \ref{assumption_nominal_controller} and \ref{assumption_high_relative_degree} are satisfied.
With conditions \eqref{eq_ubarg_bardelta}--\eqref{eq_chosen_of_bc}, \eqref{eq_bound_f_x_1}--\eqref{eq_def_k_i} and 
\begin{align}
\gamma_{i,j} < \id \text{~and~} \gamma_{i,1}<\gamma_{1,2}^{-1}, ~~~ i,j=1,\ldots,m,i\neq j  \label{eq_smallgain_kappa_i}
\end{align}
satisfied by appropriately choosing safety parameters $\theta$, $\alpha$, $c_j$ and $\alpha_j$ for $j=1,\ldots,n_c$, reshaping parameters $c_A$, $A_L$, $k_{\phi}$ and tracking parameters $\kappa_i$ for $i=2,\ldots,m$, the safety control objective \eqref{eq_objective_highorder} is achievable with 
\begin{align}
\breve{c} &= \max\{c_1,\ldots,c_{n_c}, \gamma^{-1}_{1,2}(\max_{i=2,\ldots,m}\gamma_{i,x_1}(c_{x_1})), \notag \\
&~~~\gamma^{-1}_{1,2}(\max_{i=2,\ldots,m}\gamma_{i,\rho_0}(\bar{\rho}_0))\}, \label{eq_choose_of_breve_c}\\
\breve{c}_i &= \max\{3\gamma_{x^*_2,V}(\breve{c}), 3\gamma_{1,2}^{-1}(\breve{c}), 3\bar{\rho}_0, \notag \\
&~~~2\max_{j=3,\ldots,i}\{\frac{\ubar{g}}{\ubar{g}-\bar{\delta}}\kappa_{j-1}\circ\gamma_{1,j-1}^{-1}(\breve{c}), \gamma_{1,j}^{-1}(\breve{c})\} \}, \label{eq_choose_of_breve_ci}
\end{align}
and $\breve{\beta},\breve{\beta}_2,\ldots,\breve{\beta}_m\in\mathcal{KL}$, where $\gamma_{1,i}=\gamma_{1,2}\circ\cdots\gamma_{i-1,i}$ for $i=2,\ldots,m$.
\end{theorem}

The proof of Theorem \ref{theorem_small_gain} is postponed to Section \ref{subsection_proof_mainresult}.

When the plant \eqref{eq_plant_dynamics} is reduced to a chain of integrators, by choosing $\kappa_i(s)=K_is$ with $K_i>0$ for $i=2,\ldots,m$, direct calculation yields the following properties: 
\begin{enumerate}
\item Assumption \ref{assumption_high_relative_degree} is satisfied with $\bar{f}_{x_i}=0$ for $i=1,\ldots,m$, $\ubar{g}=1$, $\bar{g}=1$, and $\bar{\delta}=0$;
\item Conditions \eqref{eq_alpha_j_definition}--\eqref{eq_chosen_of_bc} of Theorem \ref{theorem_small_gain} can be satisfied by choosing $\alpha$ as a linear function with a suitable slope;
\item Conditions \eqref{eq_bound_f_x_1} and \eqref{eq_def_k_i} of Theorem \ref{theorem_small_gain} are satisfied with $c_{x_1} = 0$ and $k_i = 2K_i$;
\item The functions defined in \eqref{eq_tracking_gains} are in the following specific forms: 
\begin{subequations}\label{eq_gains_for_integrators}
\begin{align}
\!\alpha_{i,x_1}(s) &= 0, ~~~~~~~~~~~~ \alpha_{i,\rho_0}(s) =\bar{k}_{1,i-1}s, \\
\alpha_{i,1}(s) &= \bar{k}_{1,i-1}\gamma_{x^*_2,V}(s), \\
\alpha_{i,j}(s) &= (\bar{k}_{j,i-1}K_j+\bar{k}_{j-1,i-1})s, \\
\alpha_{i,i}(s) &= k_{i-1}s,
\end{align}
\end{subequations}
for $j=2,\ldots,i-1$, where $\bar{k}_{j,i}=\Pi^i_{p=j} 2K_p$ for $j=2,\ldots,i$, and $\bar{k}_{1,i}=k_1\bar{k}_{2,i}$;
\item Conditions \eqref{eq_alpha_i_condition} and \eqref{eq_smallgain_kappa_i} of Theorem \ref{theorem_small_gain} can be satisfied by choosing
\begin{subequations}
\begin{align}
\gamma_{i,i+1}(s) &= s/\tau, ~~~ \gamma_{1,\rho_0}(s) = s/\tau, \\
\gamma_{i,1}(s) &= \gamma_{1,2}(s)/\tau \\
\gamma_{i,j}(s) &= s/\tau, ~~~j=1,\ldots,i-1,
\end{align}    
\end{subequations}
and selecting $K_i$ such that
\begin{align}
K_is &\ge \theta s +\tau s + \bar{k}_{1,i-1}\tau s + \bar{k}_{1,i-1}\gamma_{x^*_2,V}\circ\gamma_{1,2}^{-1}(\tau s) + \notag \\
& ~~~\sum^{i-1}_{j=2}(\bar{k}_{j,i-1}K_j+\bar{k}_{j-1,i-1})\tau s + 2K_{i-1}s, \label{eq_parameters_chosen_integrators}
\end{align}
for all $i=2,\ldots,m$, where $\theta>0$ and $\tau>1$.
\end{enumerate}

As a direct result of Theorem \ref{theorem_small_gain}, we have

\begin{corollary}\label{corollary_integrators}
Consider a plant modeled by a chain of $m$ integrators, the controller \eqref{eq_safety_controller} and \eqref{eq_reference_tracking_control_law} and the certificate functions $V_1,\ldots, V_{n_c}$. Suppose that Assumptions \ref{assumption_certificate_functions} and \ref{assumption_nominal_controller} are satisfied. Then, the safety control objective \eqref{eq_objective_highorder} is achievable by appropriately choosing $\alpha$, $\theta$, $c_j$ $\alpha_j$ $c_A$, $A_L$, $k_{\phi}$ and $\kappa_i$ for $j=1,\ldots,n_c$ and $i=2,\ldots,m$.
\end{corollary}

\begin{remark}
When the safety constraints are removed, the above results are consistent with previously developed results in nonlinear stabilization \cite{Khalil-book-2002, Sepulchre-Jankovic-Kokotovic-Book-1997, Jiang-Mareels-TAC-1997}.
\end{remark}

\subsection{Proof of Theorem \ref{theorem_small_gain}} \label{subsection_proof_mainresult}

Recall condition \eqref{eq_smallgain_kappa_i}.
By using \cite[Lemma C.1]{Liu-Jiang-Hill-Book-2018}, there exist $\hat{\gamma}_{i,i+1}\in\mathcal{K}_{\infty}$ for $i=1,\ldots,m-1$ and $\hat{\gamma}_{\bar{i},\ubar{i}}\in\mathcal{K}_{\infty}$ for $\ubar{i},\bar{i}=1,\ldots,m$, $\ubar{i}<\bar{i}$, which are continuously differentiable on $(0,\infty)$ and satisfy $\hat{\gamma}_{i,i+1}>\gamma_{i,i+1}$ for $i=1,\ldots,m-1$, $\hat{\gamma}_{\bar{i},\ubar{i}}>\gamma_{\bar{i},\ubar{i}}$ and 
\begin{align}
\hat{\gamma}_{\ubar{i},\ubar{i}+1}\circ\cdots\circ\hat{\gamma}_{\bar{i}-1,\bar{i}}\hat{\gamma}_{\bar{i},\ubar{i}} < \id,
\label{eq_small_gain_condition2}
\end{align}
for $\ubar{i},\bar{i}=1,\ldots,m$, $\ubar{i}<\bar{i}$.

Along the trajectories of the closed-loop system composed of \eqref{eq_highorder_model}, \eqref{eq_safety_controller} and \eqref{eq_reference_tracking_control_law}, denote
\begin{align}
W_i(t) = \max\{W_{i-1}(t), \hat{\gamma}_{1,i}(|\tilde{x}_i(t)|)\}, \label{eq_def_barVi}
\end{align}
for $i=2,\ldots,m$, where $W_1(t)=\breve{V}(x_1(t))$ and $\hat{\gamma}_{1,i}=\hat{\gamma}_{1,2}\circ\cdots\circ\hat{\gamma}_{i-1,i}$.
Then, substituting \eqref{eq_x_tilde_final_traj} into \eqref{eq_def_barVi}, we have
\begin{align}
W_i(t) \le \max\{&\hat{\beta}_i(X_i(0), t), \hat{\gamma}_{i,x_1}( c_{x_1} ), \hat{\gamma}_{i,\rho_0}(\bar{\rho}_0),c_1,\ldots,c_{n_c}, \notag \\
&\hat{\gamma}_{1,i}\circ\gamma_{i,1}(\|W_1\|_t),\hat{\gamma}_{i,i}\circ\gamma_{i,2}(\Vert \tilde{x}_2 \Vert_t),\ldots,  \notag \\
&\hat{\gamma}_{1,i}\circ\gamma_{i,i-1}(\|\tilde{x}_{i-1}\|_t), \hat{\gamma}_{1,i}\circ\gamma_{i,i+1}(\|\tilde{x}_{i+1}\|_t)\} \notag \\
\le \max\{&\hat{\beta}_i(X_i(0), t), \hat{\gamma}_{i,x_1}( c_{x_1} ), \hat{\gamma}_{i,\rho_0}(\bar{\rho}_0),c_1,\ldots,c_{n_c}, \notag \\
&\hspace{-25pt}\hat{\gamma}_{1,i}\circ\gamma_{i,1}(\|W_i\|_t),\hat{\gamma}_{1,i}\circ\gamma_{i,2}\circ\hat{\gamma}_{1,2}^{-1}(\Vert W_i\Vert_t), \ldots, \notag \\
&\hspace{-25pt}\hat{\gamma}_{1,i}\circ\gamma_{i,i-1}\circ\hat{\gamma}_{1,i-1}^{-1}(\|W_i\|_t), \hat{\gamma}_{1,i}\circ\gamma_{i,i+1}(\|\tilde{x}_{i+1}\|_t)\} \notag \\
\le \max\{&\hat{\beta}_i'(\max\limits_{j=1,\ldots,i} [X_i(0)]_{j}, t), \hat{\gamma}_{i,x_1}( c_{x_1} ), \hat{\gamma}_{i,\rho_0}(\bar{\rho}_0),\notag \\
& c_1,\ldots,c_{n_c}, \hat{\gamma}_{1,i}\circ\gamma_{i,i+1}(\|\tilde{x}_{i+1}\|_t)\}, \label{eq_barVi_traj}
\end{align}
for all $i=2,\ldots,m$, where $X_i(t) = [\breve{V}(x_1(t));|\tilde{x}_2(t)|;\ldots;|\tilde{x}_i(t)|]$, $\hat{\beta}_i'\in\mathcal{KL}$ and $\hat{\beta}_i:\mathbb{R}_+^{i}\times\mathbb{R}_+\rightarrow\mathbb{R}_+$, $\hat{\gamma}_{i,x_1}\in\mathcal{K}$ and $\hat{\gamma}_{i,x_1}\in\mathcal{K}$ are defined by
\begin{subequations}
\begin{align}
\hat{\beta}_i(X_i,t) &= \max\{\beta_1([X_i]_1, t), \hat{\gamma}_{1,2}\circ \beta_2([X_i]_2,t),\ldots, \notag \\
&\hspace{90pt}\hat{\gamma}_{1,i}\circ \beta_i([X_i]_i,t)\}, \\
\hat{\gamma}_{i,x_1}(s) &= \max\{\hat{\gamma}_{1,2}\circ\gamma_{2,x_1}(s),\!\ldots,\!\hat{\gamma}_{1,i}\circ\gamma_{i,x_1}(s)\}, \\
\hat{\gamma}_{i,\rho_0}(s) &= \max\{\hat{\gamma}_{1,2}\circ\gamma_{2,\rho_0}(s),\!\ldots,\!\hat{\gamma}_{1,i}\circ\gamma_{i,\rho_0}(s)\},
\end{align}
\end{subequations}
for $s\in\mathbb{R}_+$ and $t\ge 0$, and we have substituted \eqref{eq_ISpS_high_relative_degree} and \eqref{eq_x_tilde_final_traj} into \eqref{eq_def_barVi} for the first inequality, used $|\tilde{x}_i|\le \hat{\gamma}_{1,i}^{-1}(\bar{V})$ for the second inequality, and used the small-gain condition \eqref{eq_small_gain_condition2} and \cite[Lemma D.1]{Liu-Jiang-Hill-Book-2018} for the last inequality.

By combining \eqref{eq_smallgain_kappa_i} and \eqref{eq_small_gain_condition2}, it can be checked that 
\begin{align}
\hat{\gamma}_{1,i} &< \hat{\gamma}_{1,2} < \gamma_{2,1}^{-1},\\
\hat{\gamma}_{m,x_1}(s) &\le \gamma^{-1}_{1,2}(\max_{i=2,\ldots,m}\gamma_{i,x_1}(s)), \\
\hat{\gamma}_{m,\rho_0}(s) &\le \gamma^{-1}_{1,2}(\max_{i=2,\ldots,m}\gamma_{i,\rho_0}(s)).
\end{align}
Then, from \eqref{eq_barVi_traj}, we have
\begin{align}
\breve{V}(x_1(t)) &\le W_m(t) \notag \\
&\le \max\{\hat{\beta}'_m(\max\limits_{i=1,\ldots,m} [X_m(0)]_{i}, t),~ \breve{c}\}, \label{eq_safty_high_order_xtilde} 
\end{align}
holds for all $t\in\mathcal{I}$, where $\breve{c}$ is defined by \eqref{eq_choose_of_breve_c}.

For $i=3,\ldots,m$, direct calculation yields that 
\begin{align}
|x_2| &\le |x^*_2| + |\tilde{x}_2| \le \gamma_{x^*_2,V}\big(\breve{V}(x_1)\big) + \bar{\rho}_0 + |\tilde{x}_2|  \notag \\
&\le 3\max\{\gamma_{x^*_2,V}(W_m), \hat{\gamma}_{1,2}^{-1}(W_m), \bar{\rho}_0\}, \label{eq_x2_traj} \\
|x_i| &\le |x_i^*|+|\tilde{x}_i| \le \frac{\ubar{g}}{\ubar{g}-\bar{\delta}}\kappa_{i-1}(|\tilde{x}_{i-1}|) + |\tilde{x}_i| \notag \\
&\le 2\max\{\frac{\ubar{g}}{\ubar{g}-\bar{\delta}}\kappa_{i-1}\circ\hat{\gamma}_{1,i-1}^{-1}(W_m), \hat{\gamma}_{1,i}^{-1}(W_m)\} \label{eq_xi_traj}
\end{align}
This, together with \eqref{eq_barVi_traj} and $\hat{\gamma}_{1,i}^{-1}<\gamma_{1,i}^{-1}$, implies that
\begin{align}
|x_i(t)| &\le \max\{\breve{\beta}'_i(\max\limits_{i=1,\ldots,m} [X_m(0)]_{i},t),\breve{c}_i\}, \label{eq_xi_boundness_xtilde}
\end{align}
holds for $t\in\mathcal{I}$, where $\breve{c}_i$ is defined by \eqref{eq_choose_of_breve_ci} and
\begin{align}
&\breve{\beta}'_i(s,t) = \max\{3\gamma_{x^*_2,V}(\hat{\beta}'_m(s,t)), 3\gamma_{1,i}^{-1}(\hat{\beta}'_m(s,t)), 2\max_{j=3,\ldots,i}\{\notag \\
&~~~\frac{\ubar{g}}{\ubar{g}-\bar{\delta}}\kappa_{j-1}\circ\gamma_{1,j-1}^{-1}(\hat{\beta}'_m(s,t)), \gamma_{1,j}^{-1}(\hat{\beta}'_m(s,t))\} \}. \label{eq_choose_of_beta_i}
\end{align}

For each $\tilde{x}_i$ with $i=3,\ldots,m$, we have
\begin{align}
|\tilde{x}_i| &\le |x_i| + |x^*_i| \le  |x_i| + \frac{\ubar{g}}{\ubar{g}-\bar{\delta}} \kappa_{i-1}(|\tilde{x}_{i-1}|) \notag \\
&\le |x_i| + \frac{\ubar{g}}{\ubar{g}-\bar{\delta}} \kappa_{i-1}(|x_{i-1}|+\cdots+\frac{\ubar{g}}{\ubar{g}-\bar{\delta}}\kappa_2(|x_2|+\notag \\
&~~~\gamma_{x^*_2,V}\big(\breve{V}(x_1)\big) + |\rho_0(x_1)|)\cdots).
\end{align}
Then, the (local) Lipschitz continuity of $\rho_0,\ldots,\rho_m$ guarantees that there exist functions $\alpha_V, \alpha_{\rho_0}, \alpha_{x_2},\ldots,\alpha_{x_m}\in\mathcal{K}$ such that 
\begin{align}
|\tilde{x}_i(0)| \le \max\{&\alpha_V(\breve{V}(x_1(0))), \alpha_{\rho_0}(|\rho_0(x_1(0))|), \notag \\
& \alpha_{x_2}(|x_2(0)|), \ldots, \alpha_{x_m}(|x_m(0)|)\}, \label{eq_xtilde_upperbound}
\end{align}
for all $i=2,\ldots,m$. 
Substituting \eqref{eq_xtilde_upperbound} into \eqref{eq_safty_high_order_xtilde} and \eqref{eq_xi_boundness_xtilde} guarantees the achievement of the control objective \eqref{eq_objective_highorder}. This ends the proof of Theorem \ref{theorem_small_gain}.

\section{Experiment: VTOL Taking Off in Narrow Spaces}
\label{section_experiment}

In this section, we consider a safety control scenario for a vertical takeoff and landing (VTOL) vehicle. Numerical simulation and physical experiments are carried out to verify the safety control method proposed in this paper.

\subsection{Dynamic Modeling of a VTOL Vehicle and Dynamic Feedback Linearization}

\begin{figure}[!ht]
\centering
\includegraphics[width=0.6\linewidth]{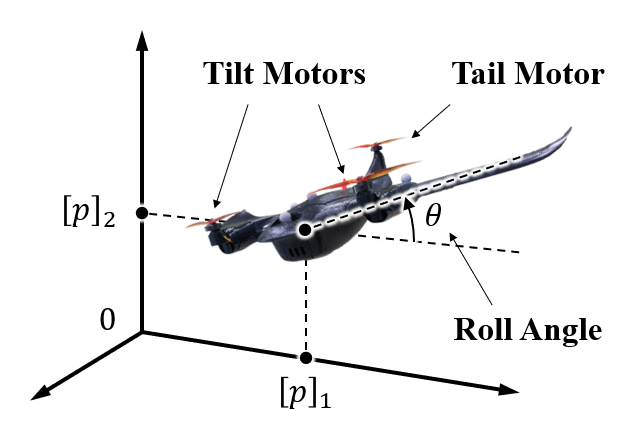}
\caption{The coordinates of the VTOL vehicle.}
\label{figure_VTOL_coordinate}
\end{figure}

Figure \ref{figure_VTOL_coordinate} shows the VTOL vehicle used in the experiment.

A simplified dynamic model of the VTOL is given by
\begin{subequations}\label{eq_VTOL_dynamics1}
\begin{align}
\ddot{p} &= \left[\begin{array}{rr} 0\\ -g\end{array}\right] + \left[\begin{array}{rr} -\sin \theta \\ \cos \theta \end{array}\right]a_1\\
\ddot{\theta} &= a_2
\end{align}    
\end{subequations}
where $p\in\mathbb{R}^2$ represents the horizontal and vertical positions of the center of mass, $\theta$ is the roll angle, and the thrust $a_1$ (directed out the bottom of the aircraft) and the angular acceleration $a_2$ (rolling moment) are used as the control inputs \cite{Fantoni-Lozano-EJC-2001}. 

As long as $a_1\neq 0$, the dynamic model can be transformed into a fourth-order integrator
\begin{align}
\displaystyle\frac{d^4}{dt^4} x_1 (t) = u(t), \label{eq_VTOL_dynamics2}
\end{align}
through the dynamic state feedback \cite{Fantoni-Lozano-EJC-2001}: 
\begin{align}
\left[\begin{array}{rr} \ddot{a}_1\\ a_2\end{array}\right]=
\left[\begin{array}{rr} \dot{\theta}^2a_1\\ -\dfrac{2\dot{\theta}\dot{a}_1}{a_1}\end{array}\right]+\left[\begin{array}{rr} -\sin\theta, &\cos \theta\\ -\dfrac{\cos\theta}{a_1}, & -\dfrac{\sin\theta}{a_1}\end{array}\right]u, \label{eq_dynamic_feedback}
\end{align}
where $x_1=p$ is the position vector, and $u\in\mathbb{R}^2$ is the snap (the fifth derivative of the position vector $p$ with respect to time).
In accordance with the constructive design procedure, we rewrite \eqref{eq_VTOL_dynamics2} as 
\begin{align}
\dot{x}_i = x_{i+1} = x^*_{i+1}+\tilde{x}_{i+1},~~~i=1,\ldots,4 \label{eq_VTOL_dyanmics3}
\end{align}
where $x^*_i$ represents the virtual control inputs, and $\tilde{x}_{i+1}$ represents the tracking-errors. Define $x^*_5=u$ and $\tilde{x}_5=0$.
Obviously, the linearized model \eqref{eq_VTOL_dyanmics3} is a chain of four integrators and satisfies Assumption \ref{assumption_high_relative_degree} with $\bar{f}_{x_1},\ldots\bar{f}_{x_4}=0$, $\ubar{g}=1$, $\bar{g}=1$ and $\bar{\delta}=0$. 
We use \eqref{eq_VTOL_dyanmics3} as the plant model in the numerical simulation.

\subsection{The Scenario for Simulation and Experiment}

Consider a scenario with $n_c$ obstacles, each represented by a line segment with ending points $o_{j_1},o_{j_2}\in\mathbb{R}^2$. Then, the safety certificate functions are defined by $V_j(x_1)=\mu(h_j(x_1))$ with 
\begin{align}
h_j(x_1)&=\begin{cases}
|x_1-o_{j_1}|-D_s,&\hspace{-5pt}\text{if~} (x_1-o_{j_1})^T(o_{j_2}-o_{j_1})<0\\
|x_1-o_{j_2}|-D_s,&\hspace{-5pt}\text{if~} (x_1-o_{j_2})^T(o_{j_1}-o_{j_2})<0\\
\frac{2S(x_1)}{|o_{j_1}-o_{j_2}|} - D_s,&\hspace{-5pt}\text{otherwise}
\end{cases}
\end{align}
where $\mu(s)=e^{-s}$, $D_s>0$ is the safe distance, $S(p)=\sqrt{l(p)(l(p)-|p-o_{j_1}|)(l(p)-|p-o_{j_2}|)(l(p)-|o_{j_1}-o_{j_2}|)}$ with $l(p)=(|p-o_{j_1}|+|p-o_{j_2}|+|o_{j_1}-o_{j_2}|)/2$.
Note that $S(p)$ represents the area of the triangle with corner points $p$, $o_{j_1}$ and $o_{j_2}$, and the area is calculated using Heron's formula \cite{Coxeter-Book-1969}. 
So, $h_j(x_1)$ basically corresponds to the distance from $x_1$ to the $j$-th line segment ($o_{j_1}, o_{j_2}$).

Taking the derivative of $h_j(x_1)$ with respect to $x_1$ yields
\begin{align}
\frac{\partial h_j(x_1)}{\partial x_1}=\begin{cases}
\frac{(x_1-o_{j_1})^T}{|x_1-o_{j_1}|},  \text{if~} (x_1-o_{j_1})^T(o_{j_2}-o_{j_1})<0 \\
\frac{(x_1-o_{j_2})^T}{|x_1-o_{j_2}|},  \text{if~} (x_1-o_{j_2})^T(o_{j_1}-o_{j_2})<0 \\
\frac{|o_{j_1}-o_{j_2}|}{4S(x_1)}(2x_1-o_{j_1}-o_{j_2})^T \\
~+\frac{|x_1-o_{j_1}|^2-|x_1-o_{j_2}|^2}{4S(x_1)|o_{j_1}-o_{j_2}|}(o_{j_1}-o_{j_2})^T, \text{otherwise}, 
\end{cases}
\end{align}
and moreover, $\left|\partial h_j(x_1)/\partial x_1\right|=1$ when $h_j(x_1)\neq -D_s$. 

In both numerical simulation and experiment, we consider $n_c=2$, $o_{1_1}=[-2.5;1.5]$, $o_{1_2}=[1.5;2.0]$, $o_{2_1}=[-2.5;0.5]$, $o_{2_2}=[2.5;0.5]$ and $D_s=0.35$. See Figure \ref{figure_simulation_trajectory} for the placement of the obstacles. Direct calculation verifies the satisfaction of Assumption \ref{assumption_certificate_functions} with $v_j=1$ for $j=1,2$. Then, using Lemma \ref{lemma_certificate_functions}, we have $\ubar{\alpha}_j(s) = \bar{\alpha}_j(s) = e^s-1$ for $j=1,2$.

\subsection{Numerical Simulation}

We consider the nominal controller 
\begin{align}
\rho_0(x_1) \equiv [0.6; 1]
\end{align}
for taking-off, which obviously satisfies Assumption \ref{assumption_nominal_controller}.

\begin{figure} [!h]
\centering
\begin{tikzpicture}[scale=0.6, transform shape]
% Safety Controller and Reference-Tracking Controller Regions
\draw[rounded corners=4pt, dashed, fill=green!5!white]  (-7.9,1.7) rectangle (-1.3,-1);
\draw[rounded corners=4pt, dashed, fill=gray!5!white]  (-0.8,1.7) rectangle (5,-1);

% Linearized VTOL Dynamics Region
\draw[rounded corners=4pt, fill=gray!20!white, line width=0.75pt] (3.4,0.6) rectangle (4.8,-0.8);
\node at (4.1,0.1) {VTOL};
\node at (4.1,-0.3) {\eqref{eq_VTOL_dynamics1}};

% Safety Controller Region
\draw[rounded corners=4pt, fill=green!20!white, line width=0.75pt] (-7.7,0.5) rectangle (-1.5,-0.7);

% Reference-Tracking Controller Region
\draw[rounded corners=4pt, fill=gray!20!white, line width=0.75pt] (-0.6,0.7) rectangle (2.4,-0.8);

% Dynamic state feedback law
\node[align=center] at (0.9,0) {Dynamic state\\ feedback law \eqref{eq_dynamic_feedback}};
\draw[-latex] (3.4,-0.3) -- (2.4,-0.3);

% Arrows and labels
\draw[-latex] (2.4,0.2) -- (3.4,0.2);
\node at (2.9,0.4) {$a_1,a_2$};
\node at (2.9,-0.6) {$\theta, \dot{\theta}$};
\node at (4.1,-1.4) {$x_1,x_2,x_3,x_4$};

% Operator Symbols
\node[align=left] at (-5,1.1) {Safety  Control Law \eqref{eq_safety_controller} and \\ Reference-Tracking Control Law \eqref{eq_reference_tracking_controller}};
\node at (1.7,1.2) {VTOL model \eqref{eq_VTOL_dynamics2}};
\node at (-4.6,-0.1) {$u=\rho_4(x_4-\rho_3(x_3-\rho_2(x_2-\rho_1(x_1))))$};
\draw[-latex] (-1.3,0.3) -- (-0.8,0.3);
\draw[-latex] (5,0.3) -- (5.3,0.3) -- (5.3,-1.8) -- (-8.2,-1.8) -- (-8.2,0.3)-- (-7.9,0.3);
\node at (-1.1,0.7) {$u$};
\end{tikzpicture}        
\caption{The block diagram of controlled VTOL.}
\label{fig_block_diagram_controlled_VTOL}
\end{figure}
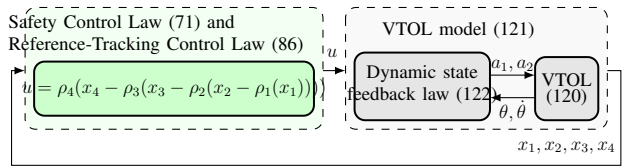

The safety controller composed of the QCQP-based safety control law \eqref{eq_safety_controller} and the tracking control law \eqref{eq_reference_tracking_control_law} is verified by the simulation. 
Figure \ref{fig_block_diagram_controlled_VTOL} shows the block diagram of controlled VTOL.

For the QCQP-based safety control law, we choose
\begin{subequations}
\label{eq_safety_controllaw_parameters}
\begin{align}
[A_c(x_1)]_{j,:} &= \frac{\partial h_j(x_1)}{\partial x_1}, &\hspace{-20pt}[c_c(x_1)]_{j}&= 0\\
[b_c(x_1)]_{j} &= k_\alpha h_j(x_1), &c_A &= \cos(\frac{2\pi}{n_l}),\\
[A_L]_{i,:} &= [\cos(\frac{2\pi i}{n_l}), \sin(\frac{2\pi i}{n_l})],\hspace{-10pt} &k_{\phi} &= 2,
\end{align}
\end{subequations}
where $k_{\alpha}=1.0$ and $i=1,\ldots, n_l$ with $n_l=11$. 
It can be verified that conditions \eqref{eq_ubarg_bardelta}, \eqref{eq_feasibility_condition} and \eqref{eq_chosen_of_bc} required by Theorem \ref{theorem_small_gain} are satisfied with $\ubar{g}=1$, $\bar{\delta}=0$, $\alpha(s)=k_{\alpha}s$, $\bar{f}_z=\bar{f}_{x}=0$, $c_j = 1.4$ for $j=1,2$, $\theta=0.001$ and $\gamma^w(s)=\gamma_{1,2}(s)=4s$. 
According to \eqref{eq_alpha_j_definition}, we choose $\bar{\alpha}_j(s)=k_{\alpha}\bar{\alpha}^{-1}_j(s)$ for $j=1,2$.

For the tracking control law, following the discussion before Corollary \ref{corollary_integrators}, we set $\kappa_i(s) = K_i s$, and choose
\begin{align}
K_2 &= 8,& K_3&= 320 & K_4&= 4.0\times 10^5. \label{eq_qp_safe_parameters}
\end{align}
It can be verified that conditions \eqref{eq_bound_f_x_1}--\eqref{eq_def_k_i} and \eqref{eq_smallgain_kappa_i} required by Theorem \ref{theorem_small_gain} are satisfied with $\bar{f}_{x_1}=0$, $k_i=2K_i$. 
Note that we use $\gamma_{x^*_2,V}(s)=k_{\alpha} s/4$, $k_1=3.49$, $\tau=1.001$ and $\theta=0.001$ for the calculation of $K_i$. 
Techniques in \cite{Wood-Zhang-JGO-1996} are used to estimate the Lipschitz constant $k_1$ of $\rho_1$.

For comparison, we also consider another set of controller parameters with the same safety control law parameters as given by \eqref{eq_safety_controllaw_parameters} and with 
\begin{align}
K_2 &= 8, &K_3&=8, &K_4&=8, 
\end{align}
corresponding to the tracking control law. With this set of controller parameters, one cannot find $\gamma_{i,1}$ for $i=2,\dots,4$ to satisfy conditions \eqref{eq_alpha_i_condition} and \eqref{eq_smallgain_kappa_i} required by Theorem \ref{theorem_small_gain} at the same time.

\begin{figure}[h]
\centering
\includegraphics[width=0.8\linewidth]{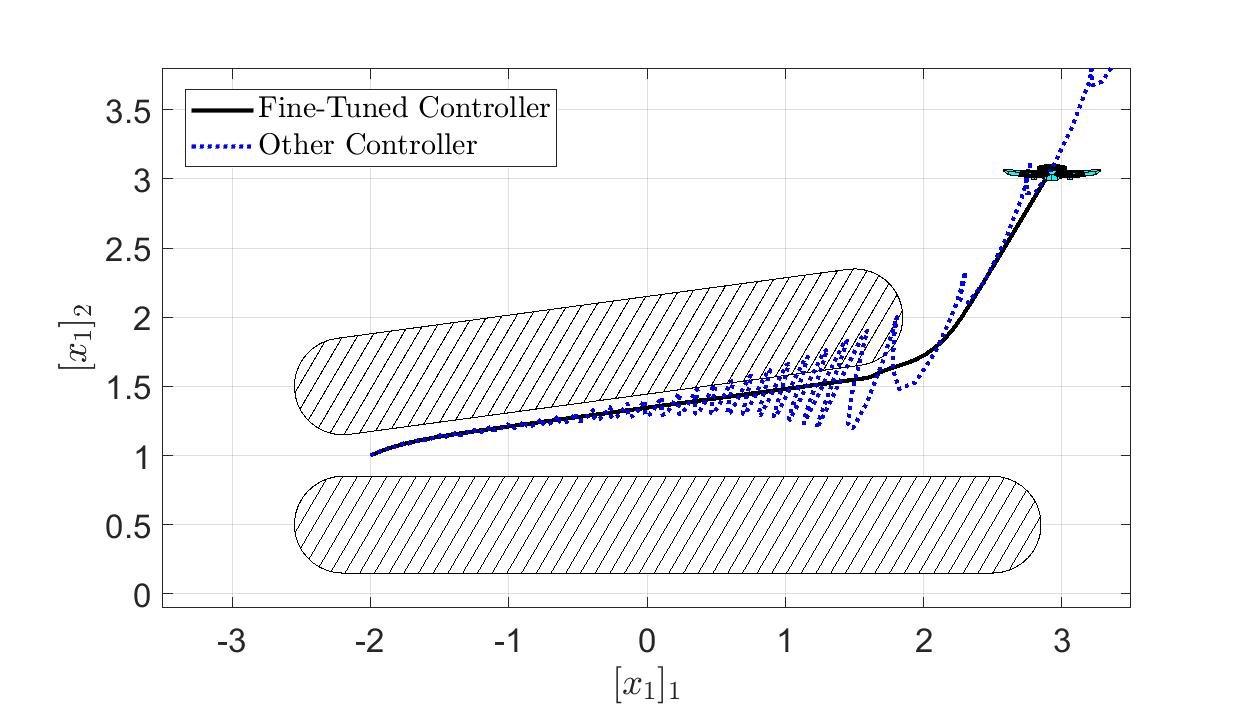}
\caption{The trajectories of the controlled VTOL with different controller parameters.}
\label{figure_simulation_trajectory}
\end{figure}

Figure \ref{figure_simulation_trajectory} shows the simulation result with $x_1(0)=[-2.0;1.0]$, and $x_i=[0;0]$ for $i=2,\ldots,4$.
The position trajectory in black corresponds to the first parameter set and maintains far away enough from the obstacles as expected, while the other trajectory violates the safety constraints. 

\subsection{Experiment}

This subsection employs experiments to further verify the validity of the proposed method.

\begin{figure}[!ht]
\centering
\includegraphics[width=0.9\linewidth]{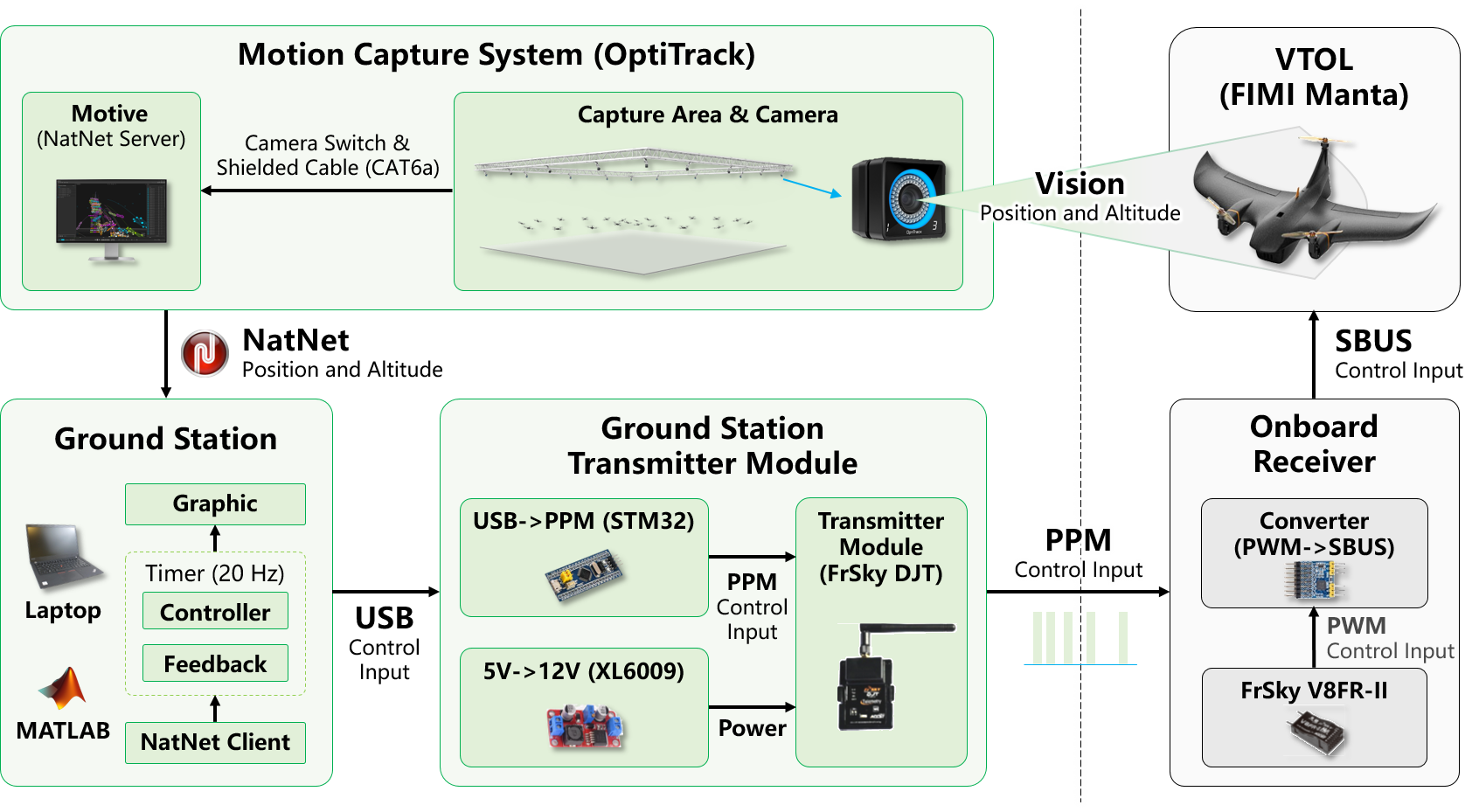}
\caption{The hardware diagram.
The plant is a FIMI Manta tiltrotor VTOL fixed-wing. 
The real-time position and attitude of the VTOL are measured by OptiTrack. 
The controller is implemented on a laptop with Intel Core i7-8565U processor. 
The communication from the laptop to the VTOL is established with a combination of a USB-to-PPM transmitter module (based on modified FrSky DJT) and a PPM-to-SBUS onboard receiver module (based on modified FrSky V8FR-II).}
\label{figure_hardware_diagram}
\end{figure}

The VTOL is already equipped with an velocity controller. 
We directly use the velocity controller instead of the designing a new one. The identified model of the velocity loop is given as follows:
\begin{align}
\dot{x}_4 = T_4(T_3(T_2(x_2^*-x_2)-x_3)-x_4)
\end{align}
where $T_2 = \diag(0.2928, 2.6711)$, $T_3=\diag(1.2231,29.7555)$ and $T_4 = \diag(4.1709,113.3872)$. We note that the identified model is in accordance with the cascade controller proposed in our paper.

In the experiment, most of the parameters are the same as those used for the numerical simulation, except $n_l$, $\rho_0(x_1)$ and $k_{\alpha}$. We choose $n_l=21$ and $\rho_0(x)\equiv[1;1]$, and try different values of $k_{\alpha}$ for the experiment. Following the analysis in Theorem \ref{theorem_small_gain} and Corollary \ref{corollary_integrators}, the value of $k_{\alpha}$ should be small enough to satisfy the conditions required by Theorem \ref{theorem_small_gain}. 

\begin{figure}[!ht]
\centering
\includegraphics[width=0.9\linewidth]{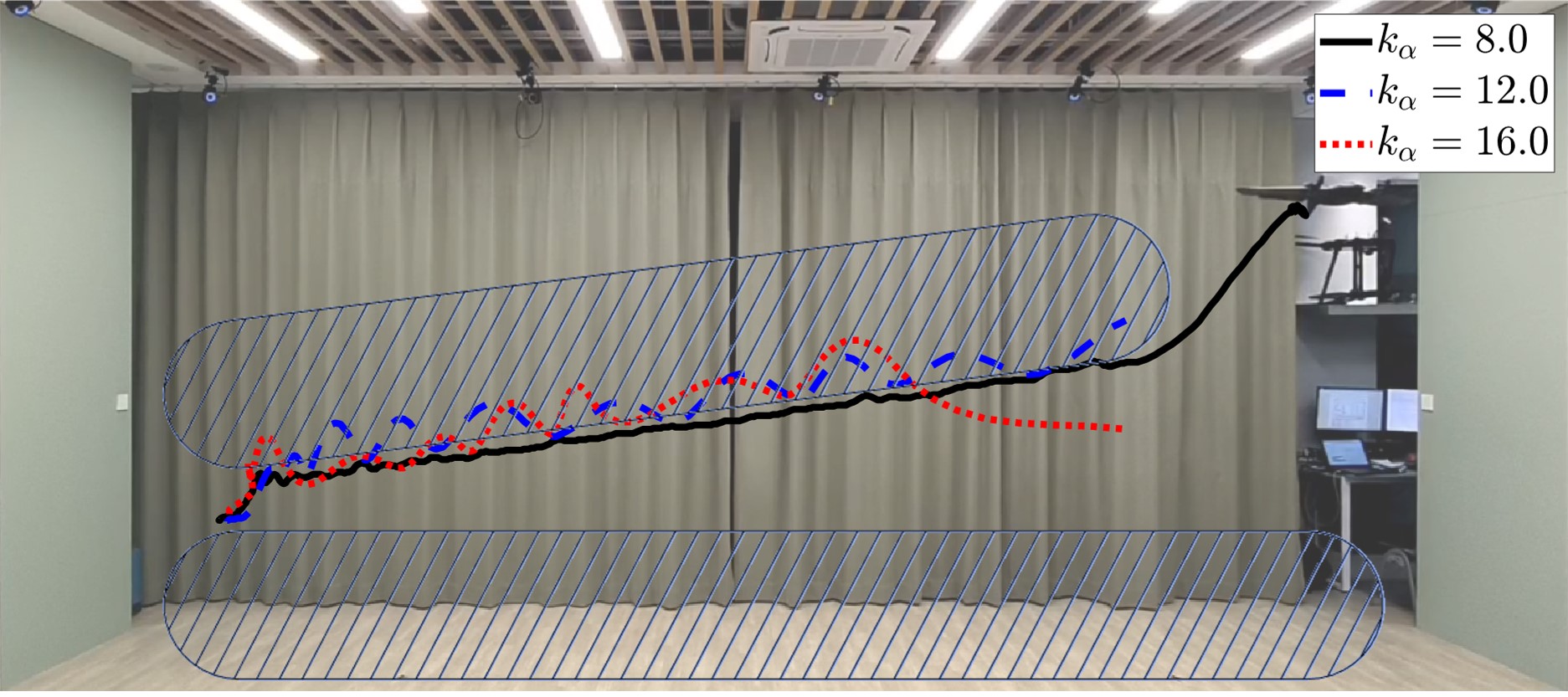}
\caption{Position trajectories of the controlled VTOL (larger values of the parameter $k_{\alpha}$ of the safety controller cause collisions).}
\label{figure_experiment_trajectory}
\end{figure}

\begin{figure}[!ht]
\centering
\includegraphics[width=0.8\linewidth]{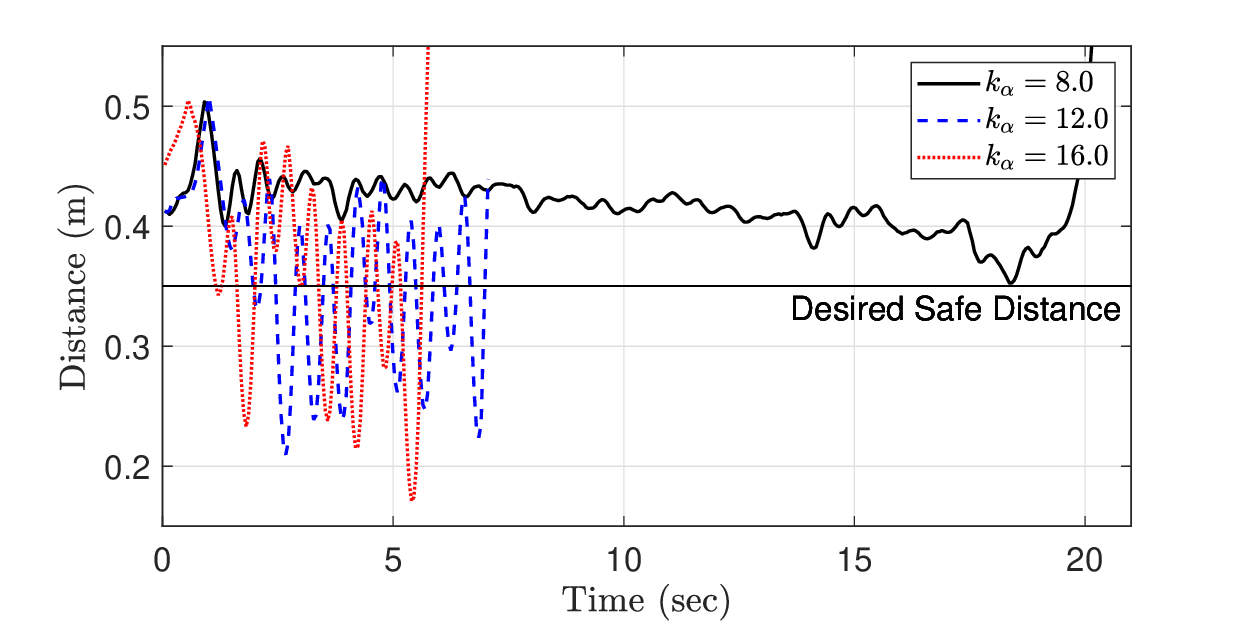}
\caption{The minimal distances between the controlled VTOL and the obstacles corresponding to different values of $k_{\alpha}$.}
\label{figure_experiment_distance_trajectory}
\end{figure}

\begin{figure}[!ht]
\centering
\includegraphics[width=0.8\linewidth]{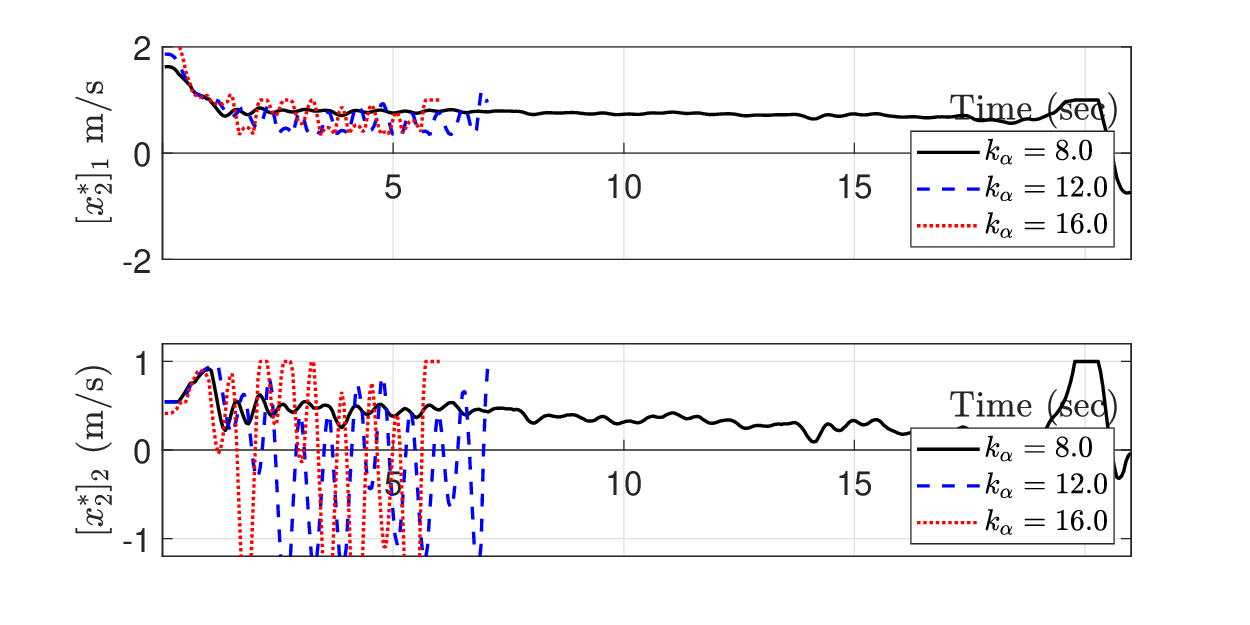}
\caption{The velocity reference signal $x^*_2$ corresponding to different values of $k_{\alpha}$.}
\label{figure_experiment_x2ref_trajectory}
\end{figure}

Figures \ref{figure_experiment_trajectory}, \ref{figure_experiment_distance_trajectory} and \ref{figure_experiment_x2ref_trajectory} show the trajectories of the controlled VTOL, the minimal distance and the velocity reference $x^*_2$ corresponding to different values of the parameter $k_{\alpha}$. 

It can be observed that the position trajectory in black corresponding to the smaller $k_{\alpha}=8.0$ maintains far away enough from the obstacles as expected, while the other two position trajectories corresponding to $k_{\alpha}=12.0$ and $k_{\alpha}=16.0$ violate the safety constraints.

\section{Conclusion}
In this paper, we have contributed a constructive solution to safety control of nonlinear uncertain systems subject to multiple safety constraints. In particular, we have first proposed a QCQP-based safety controller for nonlinear plants of relative-degree-one, with ensured feasibility under multiple safety constraints and robustness with respect to disturbances and uncertain dynamics. Then, we have developed a feasible-set reshaping technique to address the Lipschitz issue with QCQP-based safety controllers. Finally, the safety control problem for nonlinear cascade systems has been solved through a recursive design procedure and a small-gain synthesis, which transform the closed-loop system into an interconnected system and fine-tune the controller parameters to ensure safety.

We expect that the proposed new techniques will play an important role in solving the challenging safety control problems for more general systems, e.g., nonholonomic systems and underactuated systems, and control systems subject to information constraints, e.g., partial-state feedback and sampled-data feedback.

\appendices

\section{Proof of Lemma \ref{lemma_certificate_functions}} \label{appendix_proof_certificate_functions}
Define
\begin{align}
\ubar{\alpha}'_j(s) &= \inf\limits_{x\in\{x\in\mathcal{X}:D(x) \ge s\}} V_j(x)-v_j, \\
\bar{\alpha}'_j(s)  &= \sup\limits_{x\in\{x\in\mathcal{X}:D(x) \le s\}} V_j(x)-v_j,
\end{align}
for $j=1,\ldots,n_c$. The lemma can be proved by choosing $\ubar{\alpha}_j,\bar{\alpha}_j\in\mathcal{K}^e$ defined on $(-a,b)$ to admit properties \eqref{eq_assume_certificate_range} and \eqref{eq_assume_certificate_limits}, and satisfy
\begin{align}
\ubar{\alpha}_j(s)&\le\ubar{\alpha}'_j(s), &\bar{\alpha}_j(s)&\ge \bar{\alpha}'_j(s)
\end{align}
for all $s\in\{D_j(x):x\in\mathcal{X}\}$.

\section{Mathematical Derivations for Property \eqref{eq_practical_dissipation}} \label{appendix_proof_smallerfeasibleset}

Consider the terms on the left-hand side of the $j$-th constraint in $\mathcal{F}(z,x,w)$ defined by \eqref{eq_ideal_control_input_set}. Suppose that Assumptions \ref{assumption_model_relative_degree_one} and \ref{assumption_certificate_functions} are satisfied.
Then, directly applying the triangle inequality \cite[Theorem 3.3]{Apostol-book-1973} yields
\begin{subequations}\label{eq_proof_smallerfeasibleset_01}
\begin{align}
&\left|\frac{\partial V_j(x)}{\partial x}g\right|^{-1}\frac{\partial V_j(x)}{\partial x}gu =[A_c(x)]_{j,:}u,\\
&\left|\frac{\partial V_j(x)}{\partial x}g\right|^{-1}\frac{\partial V_j(x)}{\partial x}\delta(z,x)u \le \frac{\bar{\delta}}{\ubar{g}}|u|, \\
&\left|\frac{\partial V_j(x)}{\partial x}g\right|^{-1}\frac{\partial V_j(x)}{\partial x}(g+\delta(z,x))w \le (1+\frac{\bar{\delta}}{\ubar{g}})|w|,\\
&\left|\frac{\partial V_j(x)}{\partial x}g\right|^{-1}\frac{\partial V_j(x)}{\partial x} f(z,x) \le \frac{\bar{f}_z(|z|)+\bar{f}_x(|x|)}{\ubar{g}},
\end{align}
\end{subequations}
for all $z\in\mathbb{R}^{n_z}$, $x\in\mathcal{X}$ and $w\in\mathbb{R}^{n_u}$. Then, taking the derivative of $V_j$ along the trajectories of the closed-loop system composed of \eqref{eq_model_relative_degree_one} and \eqref{eq_set_valued_control_practical}, we have
\begin{align}
\nabla V_j\dot{x} &= \frac{\partial V_j(x)}{\partial x}\Big((g+\delta(z,x))(u+w)+f(z,x)\Big)\notag \\
&\le \left|\frac{\partial V_j(x)}{\partial x}g\right|\Big([A_c(x)]_{j,:}u+\frac{\bar{\delta}}{\ubar{g}}|u| + (1+\frac{\bar{\delta}}{\ubar{g}})|w|\notag \\
&~~~+\frac{1}{\ubar{g}}\bar{f}_x(|x|)+\frac{1}{\ubar{g}}\bar{f}_z(|z|)\Big)
\end{align}
where we have used the plant dynamics \eqref{eq_model_relative_degree_one} for the equality, and used \eqref{eq_proof_smallerfeasibleset_01} and the definition of $A_c$ in \eqref{eq_practical_control_input_set_A_c} for the inequality. Using $u\in\mathcal{F}_c(x)$ with $\mathcal{F}_c$ defined in \eqref{eq_practical_control_input_set} proves \eqref{eq_practical_dissipation}.

\section{A Theorem for Lipschitz Continuity of QP Solutions} \label{appendix_hager}

Here recalls \cite[Theorem 3.1]{Hager-SIAMControl-1979}, which plays a vital role in proving the Lipschitz continuity property of the QP problems with reshaped feasible sets in our paper.

\begin{theorem}\label{theorem_hager} 
Consider the QP problem defined by \eqref{eq_nonLipschitz_Optimization} with $c_c(x)\equiv 0$.
For each $x\in\mathcal{X}$, we represent the non-redundant active constraints as $\breve{A}_c(x) u= \breve{b}_c(x)$. 
Suppose that the following conditions are satisfied:
\begin{itemize}
\item $\{\rho_0(x):x\in\mathcal{X}\}, \{A_c(x):x\in\mathcal{X}\}, and \{b_c(x):x\in\mathcal{X}\}$ are convex;
\item For each $x\in\mathcal{X}$, the QP problem admits a unique solutions, denoted by $\rho_L(x)$;
\item There exists a constant $\bar{A}_c > 0$ such that $|A_c^T(x)|<\bar{A}_c$ for all $x\in\mathcal{X}$;
\item There exists a constant $\ubar{A}_c > 0$ such that for each $x\in\mathcal{X}$, $|\breve{A}_c^T(x)\lambda|\ge \ubar{A}_c|\lambda|$ for all $\lambda$.
\end{itemize}
Then, there exists a constant $L\le 1 + 2\ubar{A}_c^{-1}(1+2\bar{A}_c\max\{\ubar{A}_c^{-1}, 1\})$ such that
\begin{align}
&|\rho_L(x_1)-\rho_L(x_2)| \le \notag \\
&~~~L(|\rho_0(x_1)-\rho_0(x_2)|+|b_c(x_1)-b_c(x_2)|) \notag \\
&+ L^2(|A_c(x_1)-A_c(x_2)|+|A_c^T(x_1)-A_c^T(x_2)|)\notag \\
&~~~\times(|\rho_0(x_1)|+|\rho_0(x_2)|+|b_c(x_1)|+|b_c(x_2)|), \label{eq_varrhoL_Lipschitz}
\end{align}
for any $x_1, x_2\in \mathcal{X}$.
\end{theorem}

\section{Proof of Lemma \ref{lemma_reshape}}
\label{appendix_lemma_reshape_proof}

The lemma is proved by choosing $b_L(x)=A_L \rho_s(x)$.

For each $x\in\mathcal{X}$, $\rho_s(x)$ is the unique element of $\mathcal{U}_L(x)$.
This ensures that the set $\mathcal{U}_L(x)$ is non-empty for each $x$, and thus guarantees the validity of property $\mathcal{U}_L(x) \subseteq \mathcal{U}_c(x)$. The uniqueness is verified by using \cite[Theorem 3.1]{Davis-AJM-1954}, which states that a positive basis $\{l_1, \ldots, l_{n_l}\}$ positively spans $\mathbb{R}^{n_u}$ if and only if, for every non-zero $d$, there exists a vector in $\{l_1, \ldots, l_{n_l}\}$ such that $l_i^T d > 0$ with $i=1,\ldots,n_l$. 

The rows of $A_L$ form a positive basis, positively spanning $\mathbb{R}^{n_u}$. Suppose that there exists a $d \neq 0$ such that $A_L(\rho_s(x)+d) \leq b_L(x)$. Then, there exists a $d \neq 0$ such that $A_L d \leq 0$, which, however, this contradicts with the positive span property. This proves the uniqueness property.

Furthermore, the (local) Lipschitz continuity of $\rho_s(x)$ guarantees the existence of a (locally) Lipschitz continuous $b_L(x)$.

\section{A Technical Lemma on Selection and its Proof} \label{appendix_lemma_Lipschitz_selection}

The following lemma gives a method to find a Lipschitz selection of a set-valued map satisfying specific conditions.

\begin{lemma}\label{lemma_Lipschitz_selection}
Consider the set-valued map $\mathcal{U}_c$ in the form of \eqref{eq_cone_feasibleset} defined by $A_c$, $b_c$ and $c_c$.
Suppose that $A_c$, $b_c$ and $c_c$ are Lipschitz continuous on $\mathcal{X}$ and satisfy
\begin{align}
\frac{[b_c(x)]_j}{1+\sgn([b_c(x)]_j)[c_c(x)]_j}+\frac{[b_c(x)]_k}{1+\sgn([b_c(x)]_k)[c_c(x)]_k} \ge 0 \label{eq_Lipschitz_condition}
\end{align}
for all $j,k=1,\ldots,n_c$ and $j\neq k$.
Then, the QP problem 
\begin{subequations}\label{eq_selection}
\begin{align}
&u=\argmin_u |u|^2 \label{eq_selection_costfunction}\\
&\text{s.t.}~ [A_c(x)]_{j,:}u\le \frac{[b_c(x)]_j}{1-[c_c(x)]_j},~j=1,\ldots,n_c, \label{eq_selection_feasibleset}
\end{align}
\end{subequations} 
admits a unique solution for each $x\in\mathcal{X}$, and the function $\rho_s:\mathcal{X}\rightarrow \mathbb{R}^{n_u}$, which represents the map from $x$ to $u$ generated by the QP problem, is Lipschitz continuous and satisfies
\begin{align}
\rho_s(x)\in\mathcal{U}_c(x) \label{eq_selection_of_Uc}
\end{align}
for all $x\in\mathcal{X}$.
\end{lemma}

\begin{IEEEproof}
For $x\in\mathcal{X}$, under condition \eqref{eq_Lipschitz_condition}, the existence and Lipschitz continuity of the solution to the QP problem \eqref{eq_selection} are proved by considering the following two cases.

\begin{itemize}
\item If $\min_{j=1,\ldots,n_c}[b_c(x)]_j\ge 0$, then $0$ is the solution to the QP problem \eqref{eq_selection}, and thus
\begin{align}
\rho_s(x) = 0. \label{eq_def_rho_s_case1}
\end{align}
It can be verified that $0\in\mathcal{U}_c(x)$.

\item If $\min_{j=1,\ldots,n_c}[b_c(x)]_j<0$, then condition \eqref{eq_Lipschitz_condition} means the existence of only one element of $b_c(x)$ less than $0$.
Denote $\ubar{j}=\argmin_{j=1,\ldots,n_c}[b_c(x)]_j$. 
The active constraint is the $\ubar{j}$-th constraint of QP problem \eqref{eq_selection}, and accordingly, the solution to the QP problem \eqref{eq_selection} is
\begin{align}
\rho_s(x) = [A_c(x)]_{\ubar{j},:}^T \frac{[b_c(x)]_{\ubar{j}}}{1-[c_c(x)]_{\ubar{j}}}. \label{eq_def_rho_s_case2}
\end{align}
Under condition \eqref{eq_Lipschitz_condition}, substituting such $\rho_s(x)$ into \eqref{eq_cone_feasibleset} verifies $\rho_s(x)\in\mathcal{U}_c(x)$.
\end{itemize}

The Lipschitz continuity of $A_c$, $b_c$ and $c_c$ guarantees the Lipschitz continuity of $\rho_s$.
\end{IEEEproof}

\section{Proof of Proposition \ref{proposition_practical_reshape}} \label{appendix_proof_practical_reshape}

We first prove that $\rho_s(x) \in\mathcal{U}_L(x)$ holds for all $x\in\mathcal{X}$.
It is easy to check that $A_L\rho_s(x)\le b_L(x)$ holds for all $x\in\mathcal{X}$ by using $\phi_1$ and $\phi_2$ being nonnegative functions.

Then, the property $\mathcal{U}_L(x) \subseteq \mathcal{U}_c(x)$ holds for all $x\in\mathcal{X}$ is verified by showing that all elements of $\mathcal{U}_c(x)$ are also in $\mathcal{U}_L(x)$ for each $x\in\mathcal{X}$. 

Rewrite $\mathcal{U}_c(x)$ as
\begin{align}
\mathcal{U}_c(x)=\bigcap^{n_c}_{j=1}\mathcal{S}_j(x),
\end{align}
where 
\begin{align}
\mathcal{S}_j(x) = \{u:[A_c(x)]_{j,:}u+[c_c(x)]_j|u|\le [b_c(x)]_j \},
\end{align}
with $j=1,\ldots,n_c$. Clearly,
\begin{align}
&\mathcal{U}_L(x)  \subseteq \mathcal{U}_c(x)\Leftrightarrow \mathcal{U}_L(x) \subseteq \mathcal{S}_j(x), \forall j=1,\ldots, n_c . \label{eq_proof_condition_UL_is_a_subset_of_Uqc}
\end{align}

Rewrite $\mathcal{S}_j(x)$ as 
\begin{align}
\mathcal{S}_j(x) = \bigcap_{c\in\{c\in\mathbb{R}^{1\times n_u}:|c|\le[c_c(x)]_j\}} \mathcal{S}_c(x) 
\end{align}
where 
\begin{align}
\mathcal{S}_c(x) = \left\{u:l_c^T(x)u\le \frac{[b_c(x)]_j}{|[A_c(x)]_{j,:} + c| }\right\}
\end{align}
with $l_c^T(x) = ([A_c(x)]_{j,:} + c)/|[A_c(x)]_{j,:} + c|$.
Thus, we may prove $\mathcal{U}_L(x) \subseteq \mathcal{S}_j(x)$ by showing that $\mathcal{U}_L(x)\subseteq\mathcal{S}_c(x)$ for all $|c|\le [c_c(x)]_j$.

Using Lemma \ref{lemma_positive_basis}, for any vector $l_c$, there exist at least $n_u$ vectors $l_{c_1},\ldots,l_{c_{n_u}}$ in $\mathcal{L}_c:=\{l_i\in\{l_1,\ldots,l_{n_l}\}: l_i^Tl_c\ge c_A\}$ such that $l_c$ is a positive combination of $\{l_{c_1},\ldots,l_{c_{n_u}}\}$. Denote the constraints in $\mathcal{U}_L$ corresponding to $l_{c_1},\ldots,l_{c_{n_u}}$ as $A_{L_c}u\le b_{L_c}$. Utilizing the definition of the positive combination \cite{Davis-AJM-1954}, there exists a $\iota\in\mathbb{R}^{n_u}_+$ (with nonnegative elements) satisfying $\iota^T A_{L_c} = l_c^T$. 

For any $u_*\in\mathcal{U}_L(x)$, we have $A_{L_c}u_* \le b_{L_c}$. This together with $\iota \ge 0$, implies 
\begin{align}
&l_c^T(x) u_*=\iota^T A_{L_c}u_* \le \iota^T A_{L_c} \rho_s(x) + \iota^T A_{L_c}[A_c(x)]_{j,:}^T\notag \\
&~~~*\frac{[b_c(x)]_j-[A_c(x)]_{j,:}\rho_s(x)-[c_c(x)]_j|\rho_s(x)|}{1+[c_c(x)]_j} \notag\\
&\le l_c^T(x)u_{L}(x)\!+\! \frac{[b_c(x)]_j-[A_c(x)]_{j,:}\rho_s(x)-[c_c(x)]_j|\rho_s(x)|}{1+[c_c(x)]_j} \notag \\
&\le l_c^T(x)u_{L}(x)\!+\! \frac{[b_c(x)]_j-[A_c(x)]_{j,:}\rho_s(x)-c\rho_s(x)}{|[A_c(x)]_{j,:} + c|}\notag \\
&\le \frac{[b_c(x)]_j}{|[A_c(x)]_{j,:} + c|}
\end{align}
where we have used the definition $b_L(x)$ in \eqref{eq_def_b_L} for the first inequality, used $\iota^TA_{L_c}=l_c(x)$, $l_c^T(x)[A_c(x)]_{j,:} < 1$ and $\rho_s(x)\in\mathcal{U}_c(x)$ for the second inequality, used $|c|\le [c_c(x)]_j$ for the third inequality, and used the definition of $l_c(x)$ for the last inequality.
This proves the satisfaction of $u_*\in\mathcal{S}_c(x)$ for all $c\in\mathbb{R}^{n_u}$ satisfying $|c|\le [c_c(x)]_j$. Then we have $\mathcal{U}_L(x)\subseteq\mathcal{S}_j(x)$ for all $x\in\mathcal{X}$ and $j=1,\ldots,n_c$. 
Recalling \eqref{eq_proof_condition_UL_is_a_subset_of_Uqc}, we can prove $\mathcal{U}_L(x)\subseteq\mathcal{U}_c(x)$.

% \section*{Acknowledgment}

% The preferred spelling of the word ``acknowledgment'' in American English is 
% without an ``e'' after the ``g.'' Use the singular heading even if you have 
% many acknowledgments. Avoid expressions such as ``One of us (S.B.A.) would 
% like to thank $\ldots$ .'' Instead, write ``F. A. Author thanks $\ldots$ .'' In most 
% cases, sponsor and financial support acknowledgments are placed in the 
% unnumbered footnote on the first page, not here.

\section*{References}

\bibliographystyle{IEEEtran}
\bibliography{bpreference}

\end{document}